\documentstyle[12pt,epsf]{article}
\def\hybrid{\topmargin 0pt      \oddsidemargin 0pt
	\headheight 0pt \headsep 0pt
	\textwidth 6.25in       
        \textheight 9.5in       
	\marginparwidth .875in
	\parskip 5pt plus 1pt   \jot = 1.5ex}

\catcode`\@=11
\def\marginnote#1{}

\newcount\hour
\newcount\minute
\newtoks\amorpm
\hour=\time\divide\hour by60
\minute=\time{\multiply\hour by60 \global\advance\minute by-\hour}
\edef\standardtime{{\ifnum\hour<12 \global\amorpm={am}%
	\else\global\amorpm={pm}\advance\hour by-12 \fi
	\ifnum\hour=0 \hour=12 \fi
	\number\hour:\ifnum\minute<10 0\fi\number\minute\the\amorpm}}
\edef\militarytime{\number\hour:\ifnum\minute<10 0\fi\number\minute}

\def\draftlabel#1{{\@bsphack\if@filesw {\let\thepage\relax
   \xdef\@gtempa{\write\@auxout{\string
      \newlabel{#1}{{\@currentlabel}{\thepage}}}}}\@gtempa
   \if@nobreak \ifvmode\nobreak\fi\fi\fi\@esphack}
	\gdef\@eqnlabel{#1}}
\def\@eqnlabel{}
\def\@vacuum{}
\def\draftmarginnote#1{\marginpar{\raggedright\scriptsize\tt#1}}

\def\draft{\oddsidemargin -.5truein
	\def\@oddfoot{\sl preliminary draft \hfil
	\rm\thepage\hfil\sl\today\quad\militarytime}
	\let\@evenfoot\@oddfoot \overfullrule 3pt
	\let\label=\draftlabel
	\let\marginnote=\draftmarginnote
   \def\@eqnnum{(\theequation)\rlap{\kern\marginparsep\tt\@eqnlabel}%
\global\let\@eqnlabel\@vacuum}  }

\def\numberbysection{\@addtoreset{equation}{section}
	\def\theequation{\thesection.\arabic{equation}}}

\catcode`@=12
\relax

\def\ie{\hbox{\it i.e. }}

\def\beq{\begin{equation}}
\def\eeq{\end{equation}}
\def\bea{\begin{eqnarray}}
\def\eea{\end{eqnarray}}

\relax
\numberbysection
\hybrid

\begin{document}
\begin{titlepage}
\begin{center}
{\large\bf Lyapounov exponent and  density of states of a one-dimensional
           non-Hermitian Schr{\"o}dinger equation}\\[.3in] 
        {\bf Bernard Derrida}$^*$, {\bf Jesper Lykke Jacobsen}$^*$ {\bf and
             Reuven Zeitak}$^{\dag}$  \\ 
	 $^*${\it Laboratoire de Physique Statistique%
             \footnote{Laboratoire associ{\'e} aux Universit{\'e}s
                       Paris 6, Paris 7 et au CNRS.},\\
             Ecole Normale Sup{\'e}rieure,\\
             24 rue Lhomond,
             F-75231 Paris CEDEX 05, FRANCE. \\}
       $^{\dag}$  {\it Department of Chemical Physics, \\
             The Weizmann Institute of Science, \\
             Rehovot 76100, ISRAEL.}
\end{center}
\vskip .15in
\centerline{\bf ABSTRACT}
\begin{quotation}
{\small We calculate, using numerical methods, the Lyapounov exponent
 $\gamma(E)$ and  the density of states $\rho(E)$ at energy $E$
of a one-dimensional non-Hermitian  Schr{\"o}dinger equation
with off-diagonal disorder.
For the particular case we consider, both $\gamma(E)$ and $\rho(E)$
depend only on the modulus of $E$. 
We find a pronounced maximum of $\rho(|E|)$
at energy $E=2 / \sqrt{3}$, which seems to be linked to
the fixed point structure of an associated random map.
We show how the density of states $\rho(E)$ can be
expanded in powers of $E$.
We find 
$\rho(|E|) = \frac{1}{\pi^2} + \frac{4}{3 \pi^3} |E|^2 + \cdots$.
This expansion, which seems to be asymptotic, can be carried out 
to  an arbitrarily high order.} 
\vskip 0.5cm
\noindent
{\em Submitted to J.~Stat.~Phys.}
\end{quotation}
\end{titlepage}

\newpage

\section{Introduction}

It has been realised over the last few years that a number of
non-equilibrium problems can be described through the time evolution
of a non-Hermitian random Hamiltonian. Some noteworthy applications
include the motion of a particle in an imaginary vector
potential \cite{H-N96}, vortex line pinning in superconductors
\cite{H-N97,FZ2}, and growth models in population biology \cite{N-S98}.
The interest for non-Hermitian random Schr\"odinger equations increased
further \cite{G-K98,B-S-B97,Zee98,H98,E97,FZ,B-Z98,MBHGZ}
as it was realised that, at least in some cases, a transition from
localised to delocalised states could take place even in one
dimension.

The simplest examples considered were one-dimensional tight-binding
models for which the wave function satisfies
\beq
   e^h \psi_{n+1} +  e^{-h} \psi_{n-1} + V_n \psi_n = E \psi_n,
  \label{eq1}
\eeq
where $V_n$ is a random potential at site $n$  and $E$ is the energy.
For real $h$ and periodic boundary conditions ($n \equiv N + n$), it
was shown \cite{H-N96,G-K98} that the eigenvalues are located (for a large
system size) either on the real axis, or on lines which can be
understood from the knowledge of the Lyapounov exponent for $h=0$.

Here we consider another simple one-dimensional
non-Hermitian Schr\"odinger equation, first introduced by Feinberg and
Zee \cite {FZ2}:
\beq
  {\rm e}^{{\rm i}\theta_n} \psi_{n+1} +
  {\rm e}^{{\rm i}\chi_n} \psi_{n-1} = E \psi_n.
  \label{iter-psi}
\eeq
In this case there is no random potential $V_n$, and the disorder is
purely off-diagonal. Both $\theta_n$ and $\chi_n$ are uniformly
distributed between $0$ and $2 \pi$, and they are all independent. 

A numerical study \cite{FZ2} of the spectrum of this Schr\"odinger
equation suggested that the density of states was roughly uniform in a
circle of radius $\sim \pi/2$. We found it challenging to see whether
existing tools \cite{dyson,schmidt,wegner,DG} could be used to
calculate this spectrum. 

In the present work we mostly study the density of states of
Eq.~(\ref{iter-psi}) by calculating the associated Lyapounov exponent
and using the Thouless formula \cite{thouless} (which remains valid
for non-Hermitian one-dimensional models as shown in
Section~\ref{sec:model}). 

The calculation of the Lyapounov exponent, and consequently the density
of states, can be done either by Monte Carlo methods or by solving
numerically an integral equation for the Ricatti variable
$r_n = | \psi_n / \psi_{n-1} | $. This is done in
Section~\ref{sec:numerics}, where we also discuss the possibility of
singularities both in the probability distribution of $r_n$ and in
the density of eigenvalues $E$ of Eq.~(\ref{iter-psi}). 

In Section~\ref{sec:perturb} we develop a method to perturbatively
calculate the Lyapounov exponent and the density of states,
in powers of the energy $E$. This approach is an
extension of similar calculations done in the past, where the anomalous
behaviour of the density of states at the band centre was obtained in
the Hermitian case \cite{wegner,DG}.
Our expansion indicates that the radius of convergence is $E=0$
and that the series is asymptotic.

\section{The Lyapounov exponent and the density of states}
\label{sec:model}

An old and powerful way \cite{dyson,schmidt} of studying
one-dimensional random systems consists in rewriting recursion
formulae like Eqs.~(\ref{eq1})--(\ref{iter-psi}) in a matrix form.
Indeed, Eq.~(\ref{iter-psi}) can be recast as
\beq
 \left[ \begin{array}{c} \psi_{n+1} \\ \psi_n \end{array} \right] =
 M_n 
 \left[ \begin{array}{c} \psi_{n} \\ \psi_{n-1} \end{array} \right],
 \label{recursion}
\eeq
where $M_n$ is a random $2 \times 2$ matrix
\beq
M_n (E, \theta_n , \chi_n) = 
 \left[ \begin{array}{cc}
   E {\rm e}^{-{\rm i}\theta_n} & -{\rm e}^{{\rm i}(\chi_n-\theta_n)} \\
   1                            & 0 \end{array} \right],
 \label{matrix}
\eeq
so that for any choice of the energy $E$ and of the  wave function
$\psi_0$ and $\psi_1$ at sites $0$ and $1$, the wave function  $\psi_n$ for
 $n \geq 2$ reads
\beq
 \left[ \begin{array}{c} \psi_{n+1} \\ \psi_n \end{array} \right] =
   \prod_{p=1}^n 
 \left[ \begin{array}{cc}
   E {\rm e}^{-{\rm i}\theta_p} & -{\rm e}^{{\rm i}(\chi_p-\theta_p)} \\
   1                            & 0 \end{array} \right]  \ 
 \left[ \begin{array}{c} \psi_{1} \\ \psi_{0} \end{array} \right].
 \label{product}
\eeq
If $\psi_0$ and $\psi_1$ are chosen arbitrarily (\ie they are fixed,
and not tuned functions of $E$ and of all the $\theta_p$ and
$\chi_p$), one expects that for large $N$
\beq
 \lim_{N\to\infty} \frac{1}{N}
 \log \left| \psi_N \right| =
 \gamma(E),
 \label{gamma-def}
\eeq
where $\gamma(E)$ is the largest Lyapounov exponent of the product of
random matrices (\ref{product}).

Because the matrices  $M_n$ are invariant under the transformation
$M_n (E, \theta_n, \chi_n ) \to
 M_n (E e^{i \beta }, \theta_n + \beta , \chi_n + \beta )$
for any $\beta$, and because by shifting all the phases $\theta_n$ and
$\chi_n$ by a constant $\beta$ one gets an equally likely random sample,
it is clear that $\gamma(E)= \gamma(E e^{i \beta} )$. Thus
the Lyapounov exponent depends only on the modulus of $E$. For
similar reasons, the  average density $\rho(E)$ of eigenvalues is
invariant under the change $E \to E e^{i \beta}$:
\bea
 \gamma(E) = \gamma(|E|),  \nonumber \\
 \rho(E) = \rho(|E|).
\label{symmetry}
\eea

In fact these two quantities are related by the Thouless formula
\cite{thouless} 
\beq
\label{thouless}
 \gamma(E) = \int_{-\infty}^\infty {\rm d}E_x \,
             \int_{-\infty}^\infty {\rm d}E_y \,
             \rho(E_x + {\rm i}E_y)
             \log|E-(E_x + {\rm i}E_y)|,
\eeq
which can be derived as in the Hermitian case by the following argument:
Consider a finite system of $N$ sites with Dirichlet boundary conditions
($\psi_0= \psi_{N+1} =0$). If one fixes $\psi_0=0$ and $\psi_1=1$,
the $\psi_{N+1}$ calculated from the recursion (\ref{recursion}) is found
to be a polynomial of degree $N$ in $E$, the zeroes of which are the
$N$ eigenvalues $E_\alpha$ (for Dirichlet boundary conditions):
\beq
 \psi_{N+1}(E) = \exp \left( -{\rm i}\sum_{n=1}^N \theta_n \right)
                 \prod_{\alpha=1}^N (E - E_\alpha).
\eeq
Applying Eq.~(\ref{gamma-def}) for large $N$ leads to Eq.~(\ref{thouless}).

It turns out that for all $E \neq 0$, the Lyapounov exponent
$\gamma(E)$ is strictly positive by the Furstenberg theorem
\cite{furstenberg,pastur}. (This will in fact be confirmed by our
numerical results, and by the small-$E$ expansion of
Section~\ref{sec:perturb}.)
One can then argue that the density of states is independent of the
boundary conditions. Indeed, for general $E$, Dirichlet
($\psi_{N+1}(E)= \psi_0(E)=0$) and periodic boundary conditions 
($\psi_{N+1}(E) = \psi_1(E)$)
are very similar: Both require that $\psi_{N+1}(E) \sim e^{N
\gamma(E)}$, obtained by iterating Eq.~(\ref{product}), takes very
untypical small values at the eigenenergies. Thus, the eigenvalues
corresponding to respectively periodic and Dirichlet boundary
conditions should be very close, their distance being typically
exponentially small in $N$.

The Thouless formula (\ref{thouless}) can be inverted by using an analogy with
two-dimensional electrostatics. If we interpret $\gamma(E)$ in
Eq.~(\ref{thouless}) as the two-dimensional Coulomb potential created by
a charge density $\rho(E')$, it follows from Poisson's equation that
\beq
 \rho(E_x + {\rm i}E_y) = \frac{1}{2\pi} \left(
 \frac{\partial^2}{\partial E_x^2} + \frac{\partial^2}{ \partial E_y^2} \right)
 \gamma(E_x + {\rm i}E_y),
\eeq
or, exploiting the rotational symmetry of $\gamma(E)$ and of
$\rho(E)$, that
\beq
 \rho(|E|) = \frac{1}{2\pi} \left(
 \frac{{\rm d}^2}{{\rm d}|E|^2} + \frac{1}{|E|} \frac{\rm d}{{\rm d}|E|}
 \right) \gamma(|E|).
 \label{rad-diff}
\eeq

For $|E|>2$ the density of states vanishes. To see this,
consider an eigenfunction $\psi_n$ corresponding to an
eigenvalue $E_\alpha$. By applying the
triangular inequality to Eq.~(\ref{iter-psi}), one obtains 
$|E_\alpha \psi_n| \le |\psi_{n-1}| + |\psi_{n+1}|$. Choosing $n$ such that
$|\psi_n| = \max \big(|\psi_i|\big)_{i=1}^N$, this implies $|E_\alpha| \le 2$.
As the density of states $\rho(E)$ vanishes for $|E|>2$, by expanding
the logarithm in  Eq.~(\ref{thouless}) and by using the rotational
symmetry (\ref{symmetry}) of $\rho(E)$ one readily finds that 
\beq
 \gamma(E) = \log |E| \ \ \  \mbox{ for } \ \ \ |E|>2.
 \label{trivial}
\eeq

\begin{figure}
\begin{center}
\leavevmode
\epsfysize=200pt{\epsffile{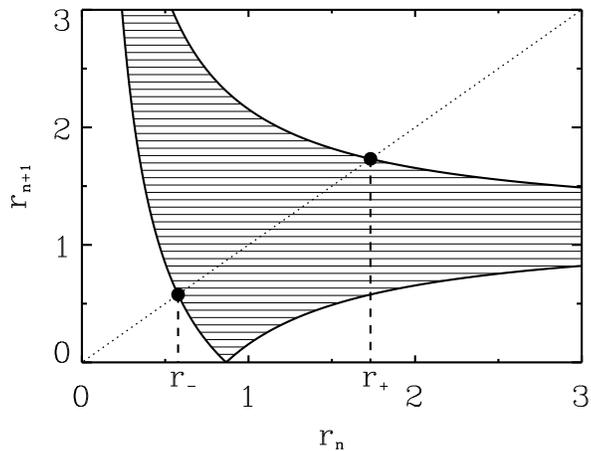}}
\end{center}
\protect\caption[5]{\label{fig:fp}Schematical illustration of the
random map (\ref{iter-r}). Given $r_n$, its iterate $r_{n+1}$ is
randomly distributed in the hatched region. The boundary constitutes
two maps, $T_+$ and $T_-$, obtained from Eq.~(\ref{iter-r}) by
choosing $\varphi_n=0$ and $\varphi_n=\pi$ respectively. For $E<2$, these maps
possess fixed points  $r_+ = T_+(r_+)$ and $r_-=T_-(r_-)$,
that are respectively attractive and repulsive. For the particular
energy $E=2/\sqrt{3}$ chosen on the figure, $T_-(r_+) = r_-$.}
\end{figure}

All our calculations which follow are based on another reformulation
of Eq.~(\ref{iter-psi}). Introducing the Ricatti variable $r_n$
\beq
  r_n = \left| \frac{\psi_n}{\psi_{n-1}} \right|
\eeq
the problem (\ref{product}) takes on the form of iterating a random map
shown in Figure~\ref{fig:fp}
\bea
  r_{n+1}   &=& \left| \frac{1}{r_n} + E {\rm e}^{{\rm i} \varphi_n}
                \right|, \label{iter-r} 
\eea
where $\varphi_n $  (which  is equal to $\chi_n - \theta_n + \pi$
modulo $2 \pi$) is uniformly distributed between $0 $ and $2\pi$. 
The Lyapounov exponent is then given by
\bea
  \gamma(E) &=& \lim_{N\to\infty} \frac1N \sum_{n=1}^N \log r_n.
  \label{mean-logr}
\eea

For large $n$ the probability distribution of $r_n$ becomes
independent of $n$, and the invariant distribution $P(r,E)$ satisfies
\beq
 P(r,E) = \int_0^{2\pi} \frac{{\rm d}\varphi}{2\pi} \,
        \int_0^{\infty} {\rm d}s \,
        P(s,E) \, \delta \left( r - \left| \frac{1}{s} +
        E {\rm e}^{{\rm i}\varphi} \right| \right).
 \label{inv-dist}
\eeq
{}From the knowledge of this invariant distribution, the calculation of
the Lyapounov exponent $\gamma(E)$ follows from Eq.~(\ref{mean-logr}),
and one has 
\beq
  \gamma(E) =  \int_0^\infty {\rm d}r \, P(r,E) \log r,
  \label{mean-logr1}
\eeq
whereas the density of states $\rho(E)$ is given by Eq.~(\ref{rad-diff}).

All the rest of the paper is devoted to the determination of the invariant
distribution $P(r,E)$ solving Eq.~(\ref{inv-dist}), and to the
calculation of the Lyapounov exponent $\gamma(E)$ as well as the density
of states $\rho(E)$.

\section{Numerical study}
\label{sec:numerics}
In this section we numerically determine the invariant distribution $P(r,E)$
solution of Eq.~(\ref{inv-dist}), the Lyapounov exponent $\gamma(E)$, and
the density of states $\rho(E)$.

According to Eq.~(\ref{rad-diff}) the computation of the density of
states requires both $\gamma(E)$ and its first two derivatives. Whilst
$\gamma(E)$ is rather easy to determine numerically, a precise
estimate of its derivatives with respect to the energy turns out to be
far more difficult. We have attacked the problem by two different methods.

\subsection{The Monte Carlo approach}

\begin{figure}
\begin{center}
\leavevmode
\epsfysize=150pt{\epsffile{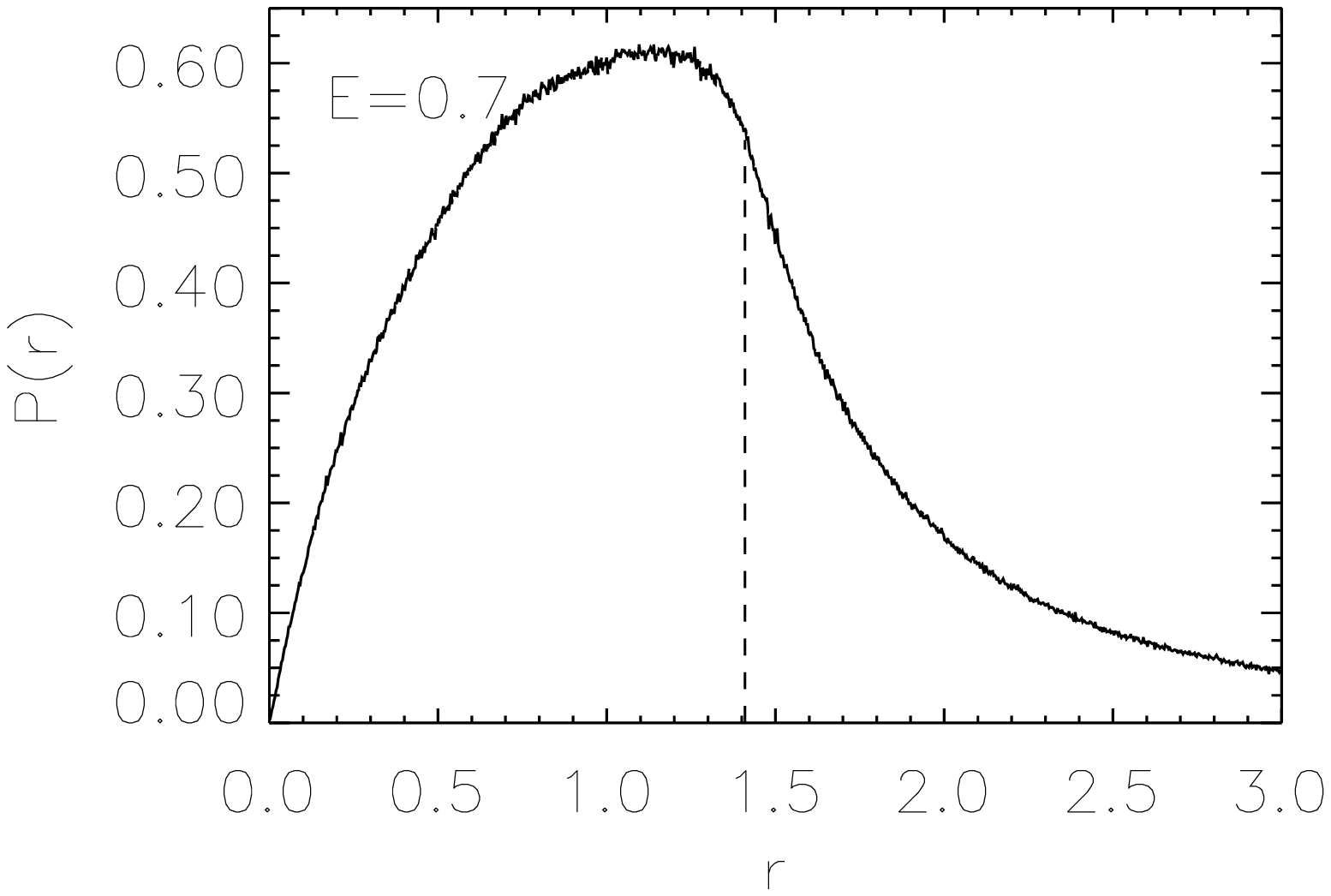}}
\epsfysize=150pt{\epsffile{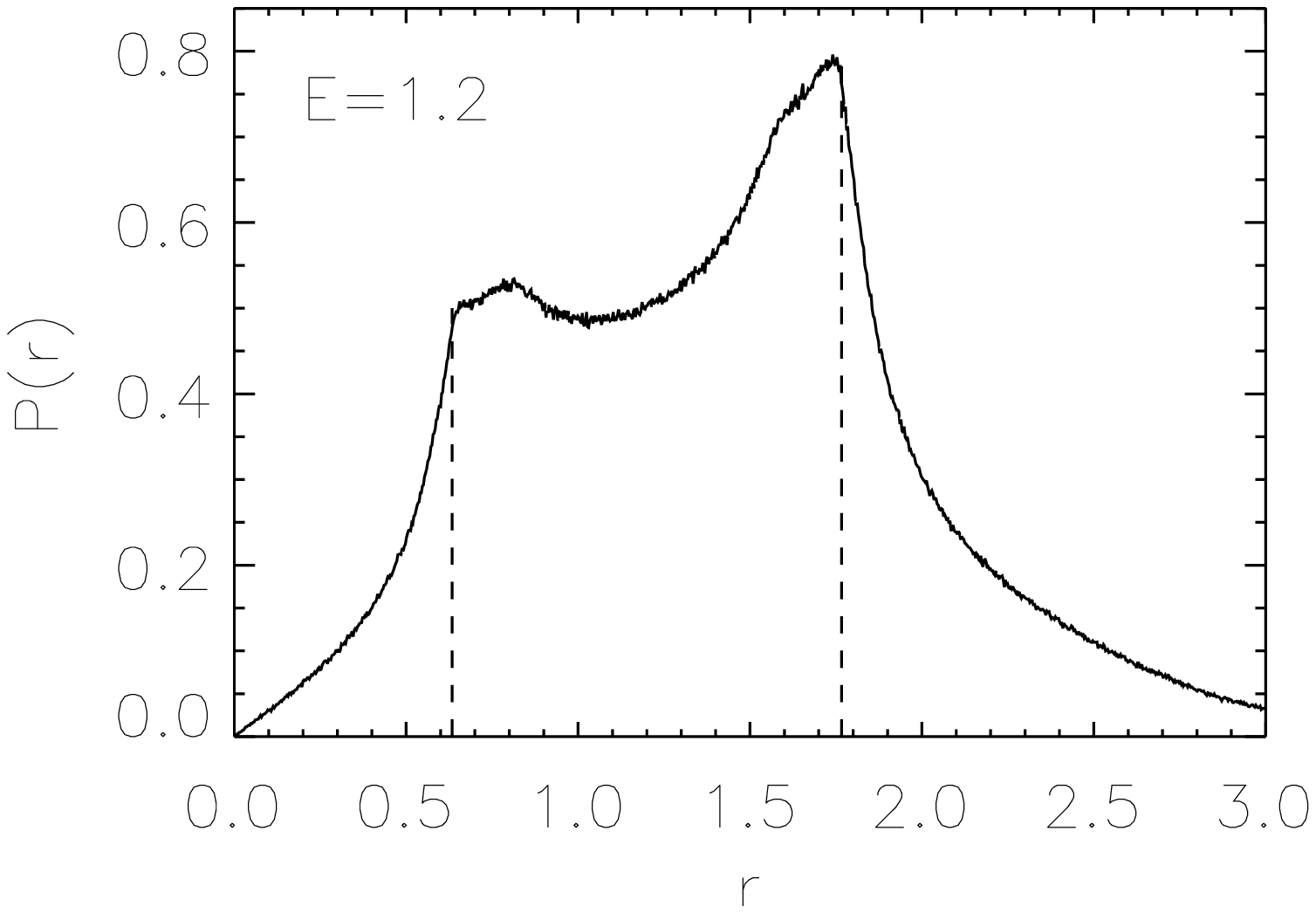}}
\epsfysize=150pt{\epsffile{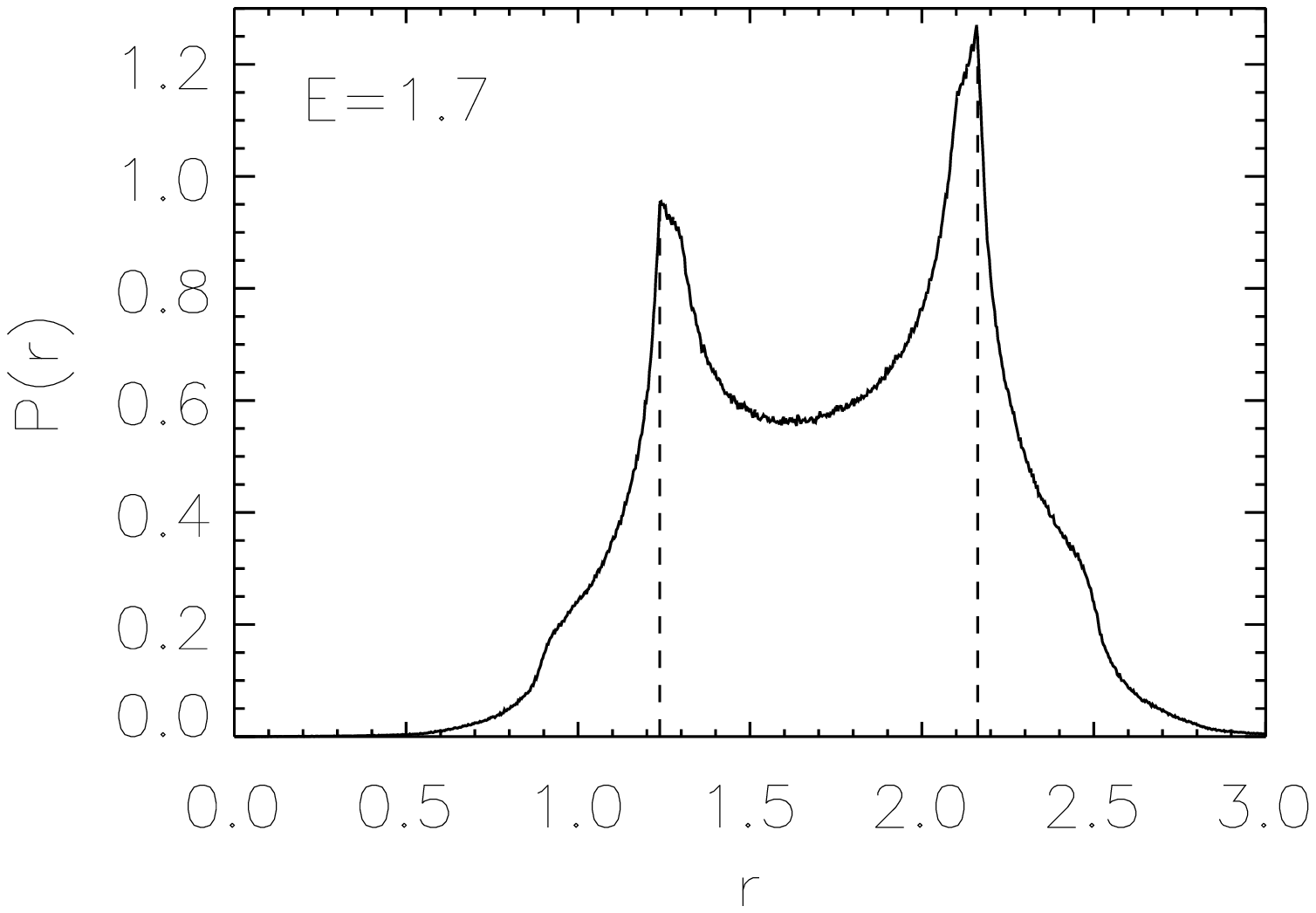}}
\epsfysize=150pt{\epsffile{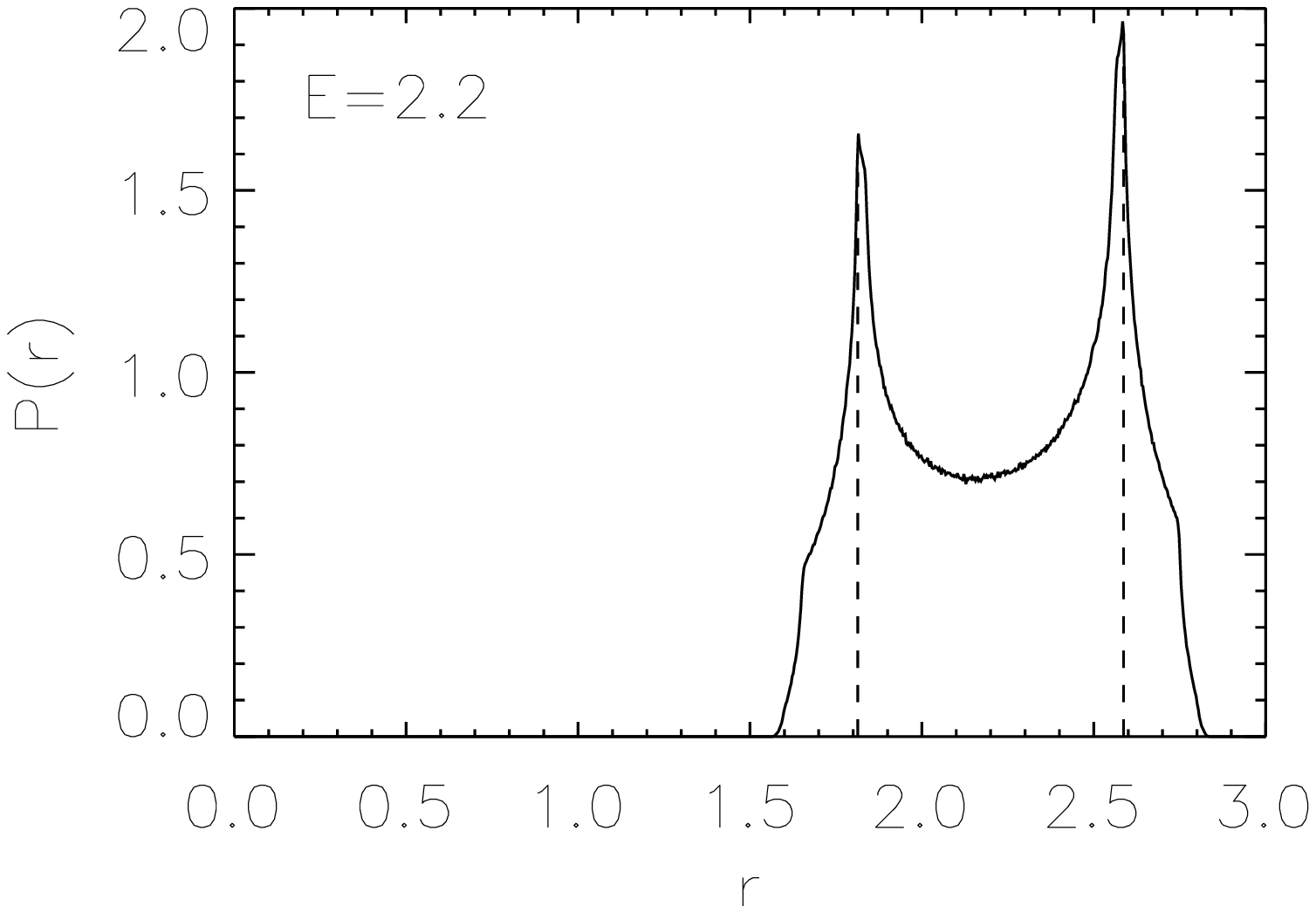}}
\end{center}
\protect\caption[5]{\label{fig:rm}Histograms of $r_n$ for energies
$E=0.7$, 1.2, 1.7 and 2.2, obtained using $N=10^7$
iterations. Apparent singularities occur at the points $r_+$ and
$T_-(r_+)$, marked by dashed lines on the figure.}
\end{figure}

This first method consists in iterating Eq.~(\ref{iter-r}) for a
large sample. Typical shapes obtained for the stationary distribution
$P(r,E)$ are shown in Figure \ref{fig:rm}.

First, one notices that depending on the value of $E$, the support of
the invariant measure $P(r,E)$ is either finite of infinite.
If one assumes that the support is an interval
$[r_{\rm min}, r_{\rm max}]$, we see from Eq.~(\ref{iter-r}) that if
$r_n \in [r_{\rm min},r_{\rm max}]$, the requirement that 
also $r_{n+1} \in [r_{\rm min},r_{\rm max}]$ gives conditions that
$r_{\rm min}$ and $r_{\rm max}$ should satisfy.
After analysing all the possible 
cases, one finds that the support of the distribution $P(r,E)$ 
is the whole positive real axis for $E < 2$, whereas for $E > 2$
the support is finite, with $ r_{\rm min}$ and $r_{\rm max}$
satisfying $r_{\rm min} = E - 1/r_{\rm min}$ and  
$r_{\rm max} = E + 1/r_{\rm min}$, that is
\beq
 r_{\rm min} = \frac{1}{2} \left( E + \sqrt{E^2 - 4} \right), \ \ \ \
 r_{\rm max} = \frac{1}{2} \left(3  E - \sqrt{E^2 - 4} \right). \ \ \ \
\eeq

Second, for large  enough $E$, there are apparent singularities at
certain values of $r$. 
These singularities seem to be remarkable points of the maps $T_+$ and
$T_-$  obtained from Eq.~(\ref{iter-r}) by choosing $\varphi_n=0$ or
$\varphi_n= \pi$, \ie
\bea
 T_+(r) &=& E + \frac{1}{r}, \nonumber \\
 T_-(r) &=& \left| E - \frac{1}{r} \right|.
\eea
For each value $r_n$, it is easy to see that $T_+(r_n)$ and $T_-(r_n)$
are the two extreme values that $r_{n+1}$ can take: 
$T_-(r_n)  \leq r_{n+1} \leq T_+(r_n)$.

The fixed point of $T_+$
\beq
  r_+ = \frac{1}{2} \left( E + \sqrt {E^2 + 4 } \right)
\eeq
coincides with the most visible singularity of $P(r,E)$ in
Figure~\ref{fig:rm}, whereas the 
iterate $T_-(r_+)$ gives the position of another very visible
singularity of $P(r,E)$. Other apparent singularities seem to be
located at the points $T_-^2 (r_+)$ and $T_+(T_-(r_+))$.

\begin{figure}
\begin{center}
\leavevmode
\epsfysize=200pt{\epsffile{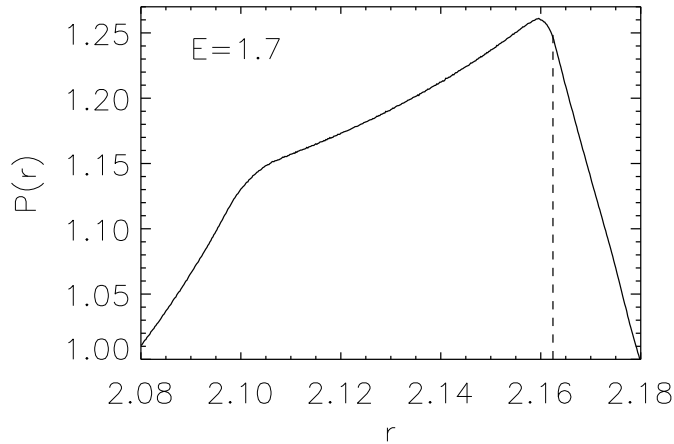}}
\end{center}
\protect\caption[5]{\label{fig:rm17}Close-up of the most visible
``singularity'' of Figure~\ref{fig:rm}.c, this time using
$N=10^{11}$ iterations.}
\end{figure}

In Appendix~\ref{app:sing}, we show that the singularities in $P(r,E)$
are, if at all present, much weaker than they 
 look on Figure~\ref{fig:rm}. This is confirmed by
Figure~\ref{fig:rm17}, which is an enlargement of
Figure~\ref{fig:rm}.c, where there is a clear rounding of the
singularity at $r_+$.

The calculation of the Lyapounov exponent is straightforward by Monte
Carlo methods.
One just iterates the random recursion (\ref{iter-r}) a large number
$N$ of times (typically $N=10^8$)
and one obtains  $\gamma(E)$ by 
\begin{equation}
 \gamma(E) \simeq {1 \over N} \sum_{n=N'}^{N+N'} \log r_n 
 \label{gmc}
\end{equation}
where $N'$ 
is a large enough number of iterations to eliminate the transient
effects due to the initial choice of $r_0$ (here we took $N'=N/10$ which is exceedingly sufficient).

For the density of states one needs Eq.~(\ref{rad-diff}) to calculate
the first two derivatives of $\gamma(E)$ with respect to $E$. A
possible approach would be to measure $\gamma(E)$ from Eq.~(\ref{gmc})
for three nearby energies, $E-\Delta E$, $E$, and $E + \Delta E$, 
and calculate numerically the first and second derivatives. 
With this method it is however hard to find a good compromise for the
choice of $\Delta E$: If $\Delta E$ is too small, one does not have
enough precision on the derivatives, and if $\Delta E$ is too large,
all the structure of the density of states is artificially smoothed. 

To avoid this difficulty we tried to iterate directly the derivatives
of $r_n$ with respect to $E$. The recursion (\ref{iter-r}) has the form
\beq
 r_{n+1} = f_0(r_n, E, \varphi_n).
 \label{f0}
\eeq
Now, if $r_n^{(1)}$ and $r_n^{(2)}$ denote the first and the second
derivatives of $r_n$ with respect to $E$, one can obtain (in a
complicated form which we do not reproduce here because we wrote it in
our programmes through a change of variables) recursion formulae for
these derivatives
\bea
 r_{n+1}^{(1)} &=& f_1(r_n,r_{n}^{(1)}, E, \varphi_n), \label{f1} \\
 r_{n+1}^{(2)} &=& f_2(r_n,r_{n}^{(1)}, r_{ n}^{(2)}, E, \varphi_n). \label{f2}
\eea
Iterating these recursions $N$ times yields an estimate for the first
two derivatives of $\gamma(E)$  
\bea
 \frac{{\rm d} \gamma(E)}{{\rm d} E}  &\simeq&
  \frac{1}{N} \sum_{n=N'}^{N+N'} \frac{r_n ^{(1)}}{r_n}, \nonumber \\
 \frac{{\rm d}^2 \gamma(E)}{{\rm d}E^2}  &\simeq&
  \frac{1}{N} \sum_{n=N'}^{N+N'} \left[\frac{r_n^{(2)}}{r_n} -
  \left( \frac{r_n^{(1)}}{r_n} \right)^2 \right]. \label{2sums}
\eea
Figure 3 shows our results for $\gamma(E)$ and $\rho(E)$ obtained by
this Monte Carlo approach.
Whilst the results for $\gamma(E)$ seem to be alright, the density of
states exhibits some irregularities. We tried to understand the origin
of these irregularities by changing the length $N$ of the sample, the
generator of random numbers, and the way in which
Eqs.~(\ref{f1})--(\ref{f2}) were parametrised in our programmes.
Although the positions of the irregularities were observed to change,
we were unable to eliminate them altogether. Nevertheless we believe
that they do have a purely numerical origin: From time to time, there
is a very small value of $r_n$ produced by the iteration of
Eqs.~(\ref{f0})--(\ref{f2}), and this gives such a huge contribution
to the sums (\ref{2sums}) that all the remaining terms become
negligible. Usually this big contribution is followed at the next step
by another huge contribution of opposite sign which more or less
compensates the first big contribution. 
These events (where $r_n$ is very small) have a dramatic
effect on the accuracy of our results, with the unfortunate
consequence that the more we iterate, the more of these events we see,
so that the more irregularities are visible.

We unfortunately could not find a satisfactory way of avoiding these
difficulties, and so we tried a completely different approach
discussed in the next subsection.

\begin{figure}
\begin{center}
\leavevmode
\epsfysize=150pt{\epsffile{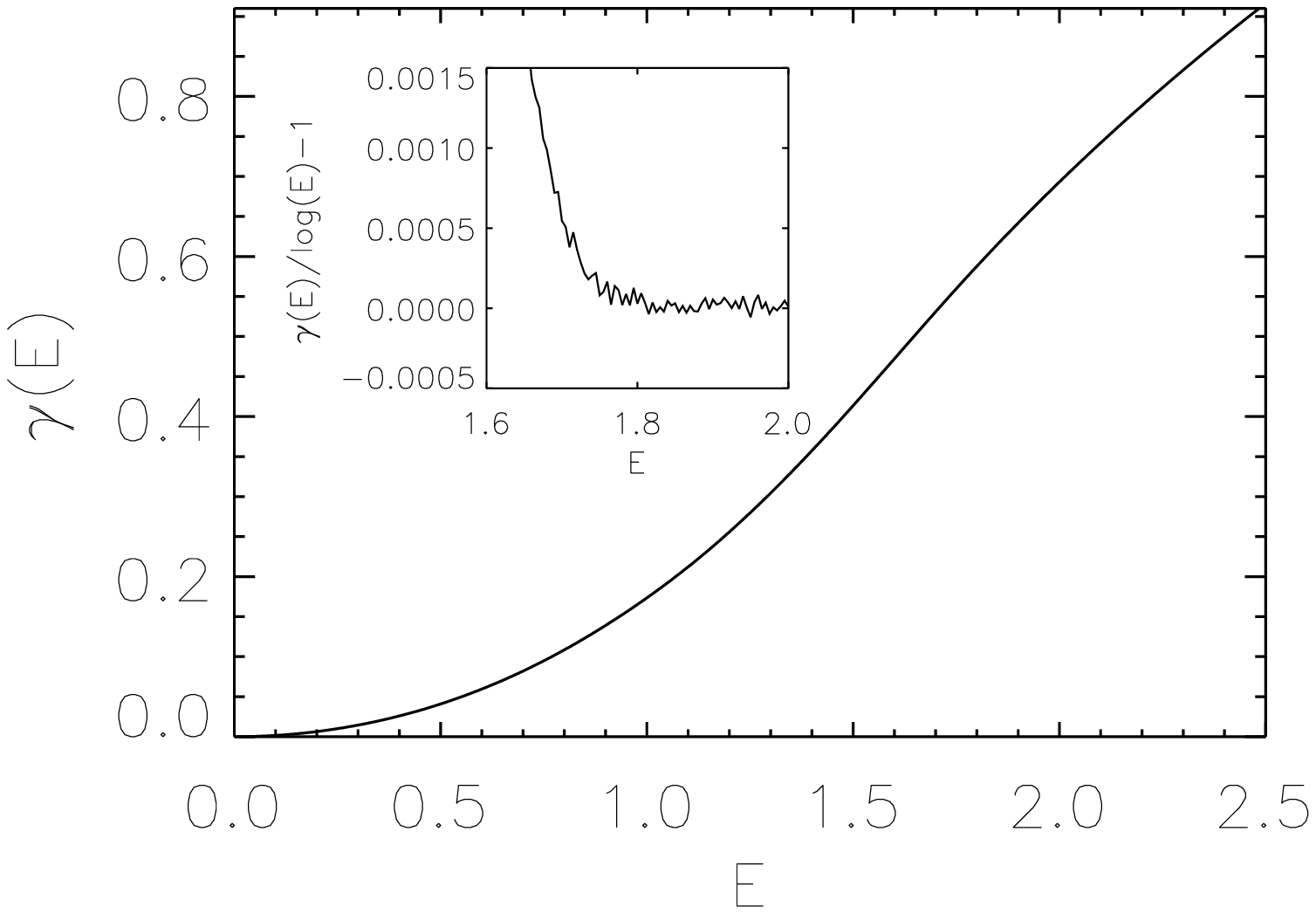}}
\epsfysize=150pt{\epsffile{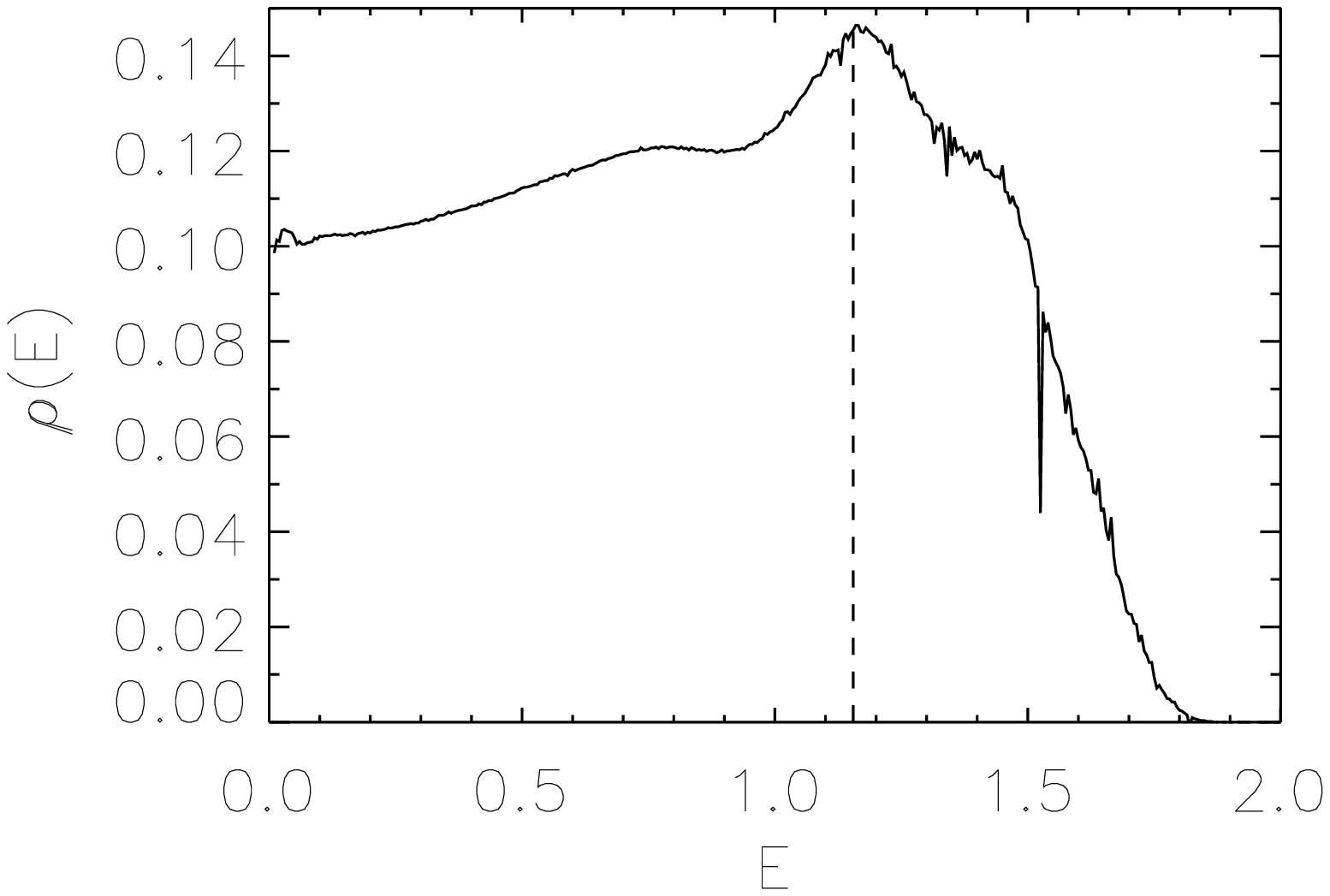}}
\end{center}
\protect\caption[5]{\label{fig:mc}Lyapunov exponent $\gamma(E)$ and
density of states $\rho(E)$ obtained by Monte Carlo after $10^8$
iterations. The inset in the left part of the figure shows the
asymptotic approach to the exact result (\ref{trivial}). In figure 4b,
the vertical dashed line indicates the energy $E=2/\sqrt{3}$.}
\end{figure}

In Figure~\ref{fig:mc}.b, we see that the density of states $\rho(E)$
exhibits a non-trivial structure with a pronounced maximum at an
energy $E \simeq 1.15$. Trying to understand the origin of this
maximum, we realised that at energy 
\beq
 E = \frac{2}{\sqrt{3}} \simeq 1.1547 \cdots
\eeq
the point $T_-(r_+)$ becomes a fixed point of $T_-$. In a more
physical language, if one considers an infinite sample for which 
\bea
\theta_n =0   \ \ {\rm and} \ \ \chi_n = \pi \ \ {\rm for} \ \   n < 0,
 \nonumber \\
\theta_n =0   \ \ {\rm and} \ \ \chi_n = 0 \ \ {\rm for} \ \ n = 0, \\
\theta_n = \pi \ \ {\rm and} \ \ \chi_n = 0 \ \ {\rm for} \ \ n > 0,
 \nonumber 
\eea
one finds that there is a bound state at energy $E=2/\sqrt{3}$
corresponding to a wave function $\psi_n \propto 3^{-|n|/2}$ in addition to
the continuous part of the spectrum at imaginary energies, $E= {\rm i}
x$ with $ -2 \leq x \leq 2$. 

We believe that the origin of the sharp maximum in $\rho(E)$ at
energy $E=2/\sqrt{3}$ is due to the existence of these bound states,
as in the  case of impurity bands for Hermitian problems
\cite{pastur,Lif}. We did not succeed, however, to determine whether the
density of states is analytic at this special energy $E= 2/\sqrt{3}$,
or whether it presents some weak non-analyticity.

Our numerical results for $\gamma(E)$ in Figure~\ref{fig:mc}.a
can be fitted for   small $E$ by an expression of the form
\beq
 \gamma(E) = k_1 E^2 + k_2 E^4 + \ldots
\eeq
with $k_1 = 0.1595 \pm 0.0005$ and $k_2 = 0.017 \pm 0.001$. 
To be more precise, we first find $k_1$ as the extrapolated
interception of $\gamma(E)/E^2$ with the ordinate. Then, subtracting
the $k_1 E^2$ term, $k_2$ can be similarly found from the residual
$E^4$ dependence. Needless to say, this procedure greatly enhances
numerical noise at each stage, and so we had to content ourselves
finding the first two terms. These are in good agreement with the results
$k_1 = \frac{1}{2\pi} \simeq 0.1592$ and
$k_2 = \frac{1}{6 \pi^2} \simeq 0.01689$ of the perturbation theory to
be developed in Section \ref{sec:perturb} (see Eq.~(\ref{lyap-exp1})).

\subsection{Discretisation of the integral equation for $P(r,E)$}

To avoid the problems of the Monte Carlo method, we tried to
numerically solve the integral equation (\ref{inv-dist}) for $P(r,E)$.
To this end the $r$-variable was discretised by taking 1000 values of
$r$ at the points $r_k=2k/(2001-2k)$ for $1 \leq k \leq 1000$. The
integral operator (\ref{inv-dist}) then becomes a
$1000 \times 1000$ matrix, and the stationary distribution can be
obtained by simply iterating a linear system.

\begin{figure}
\begin{center}
\leavevmode
\epsfysize=150pt{\epsffile{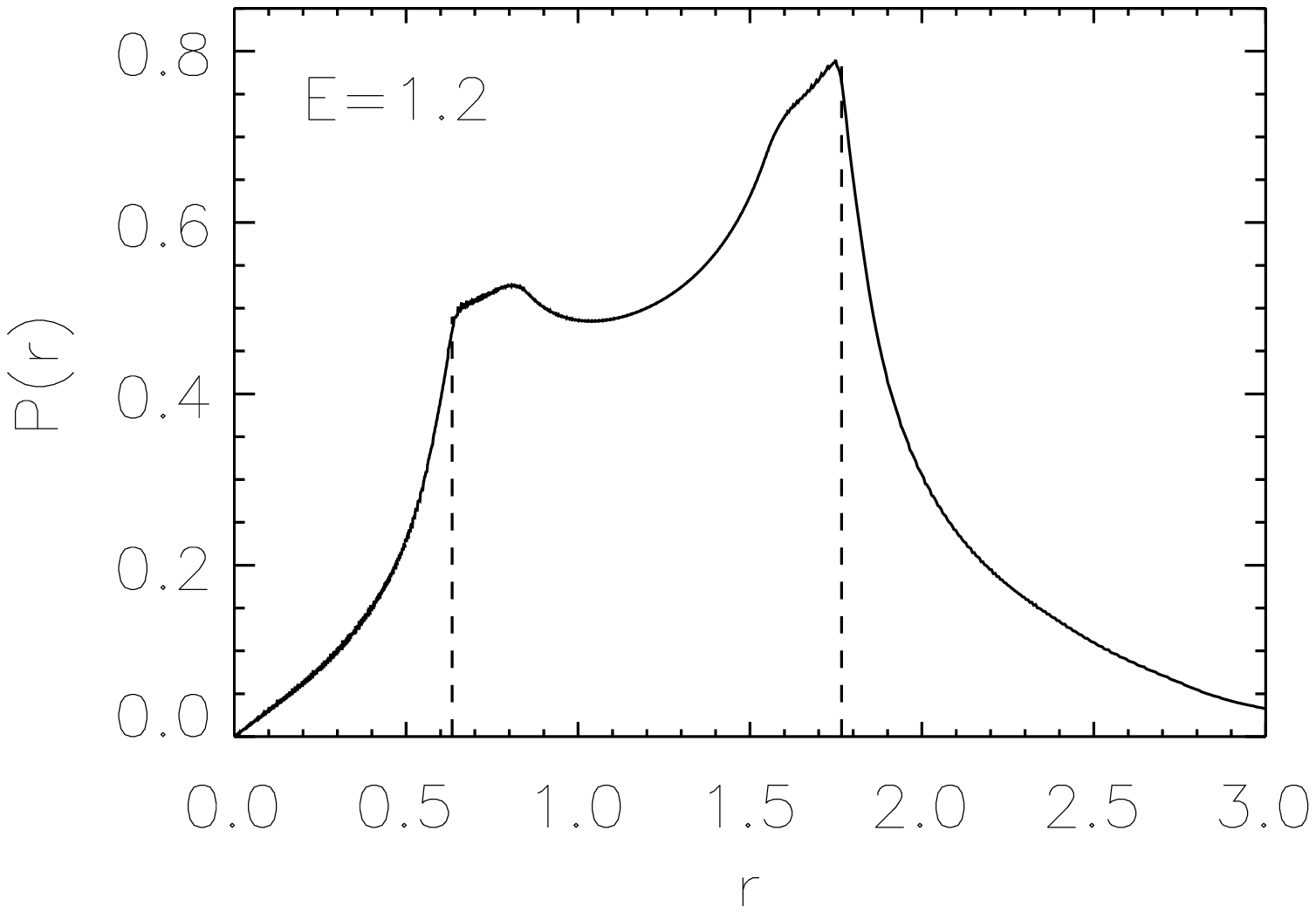}}
\epsfysize=150pt{\epsffile{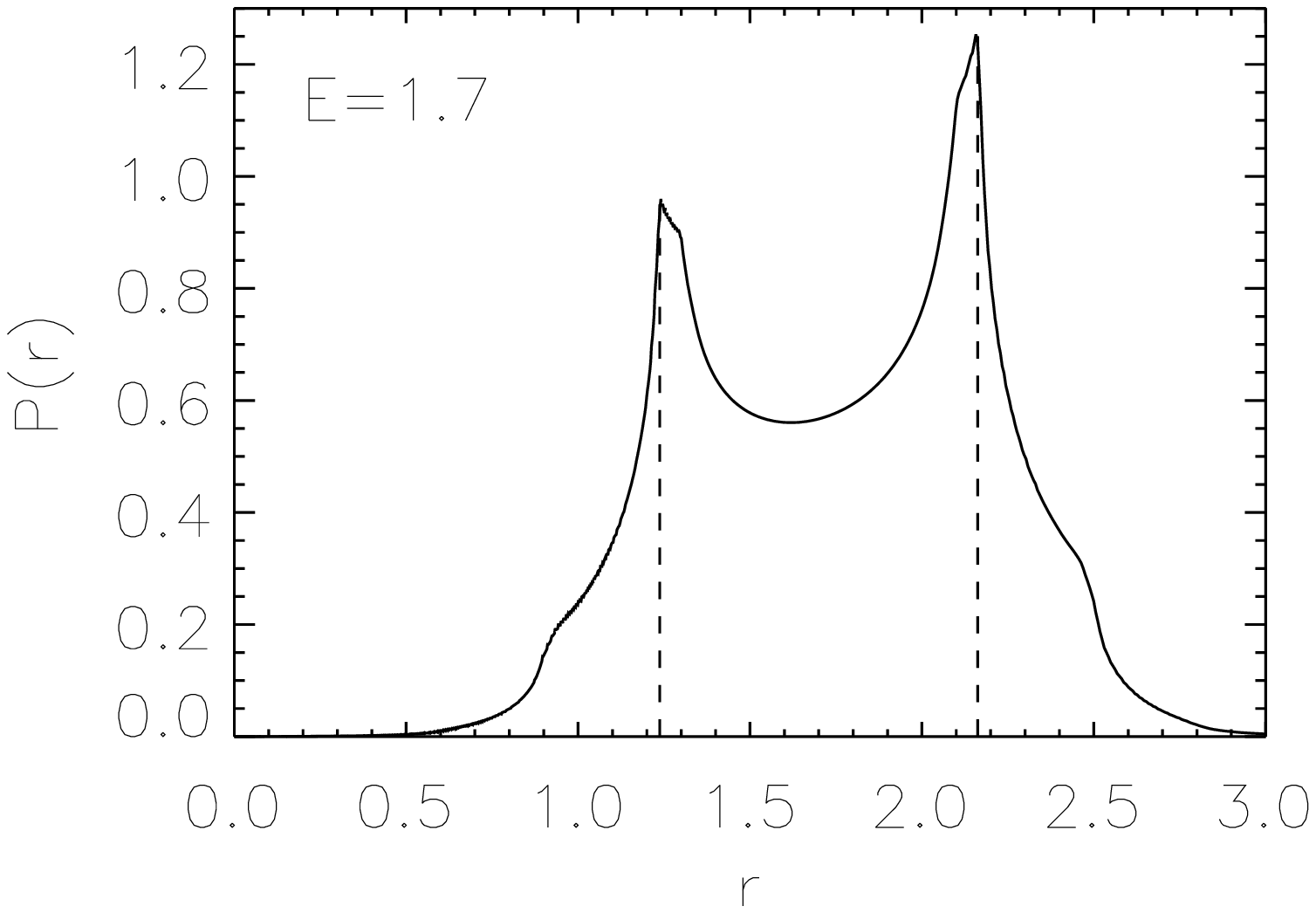}}
\end{center}
\protect\caption[5]{\label{fig:por}Invariant distributions $P(r,E)$ for
energies $E=1.2$ and 1.7, obtained by discretisation of the integral
equation (\ref{inv-dist}).}
\end{figure}

\begin{figure}
\begin{center}
\leavevmode
\epsfysize=150pt{\epsffile{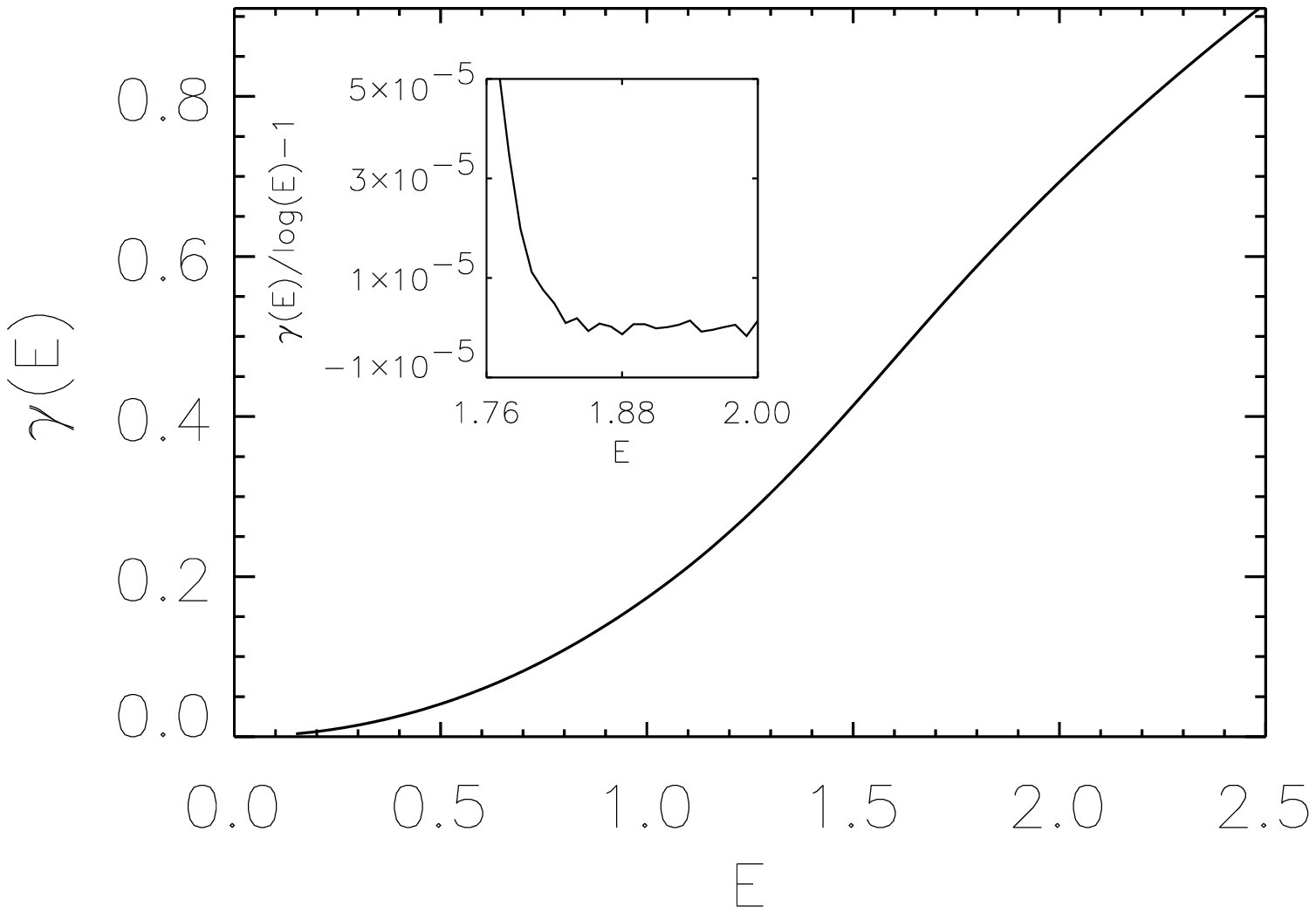}}
\epsfysize=150pt{\epsffile{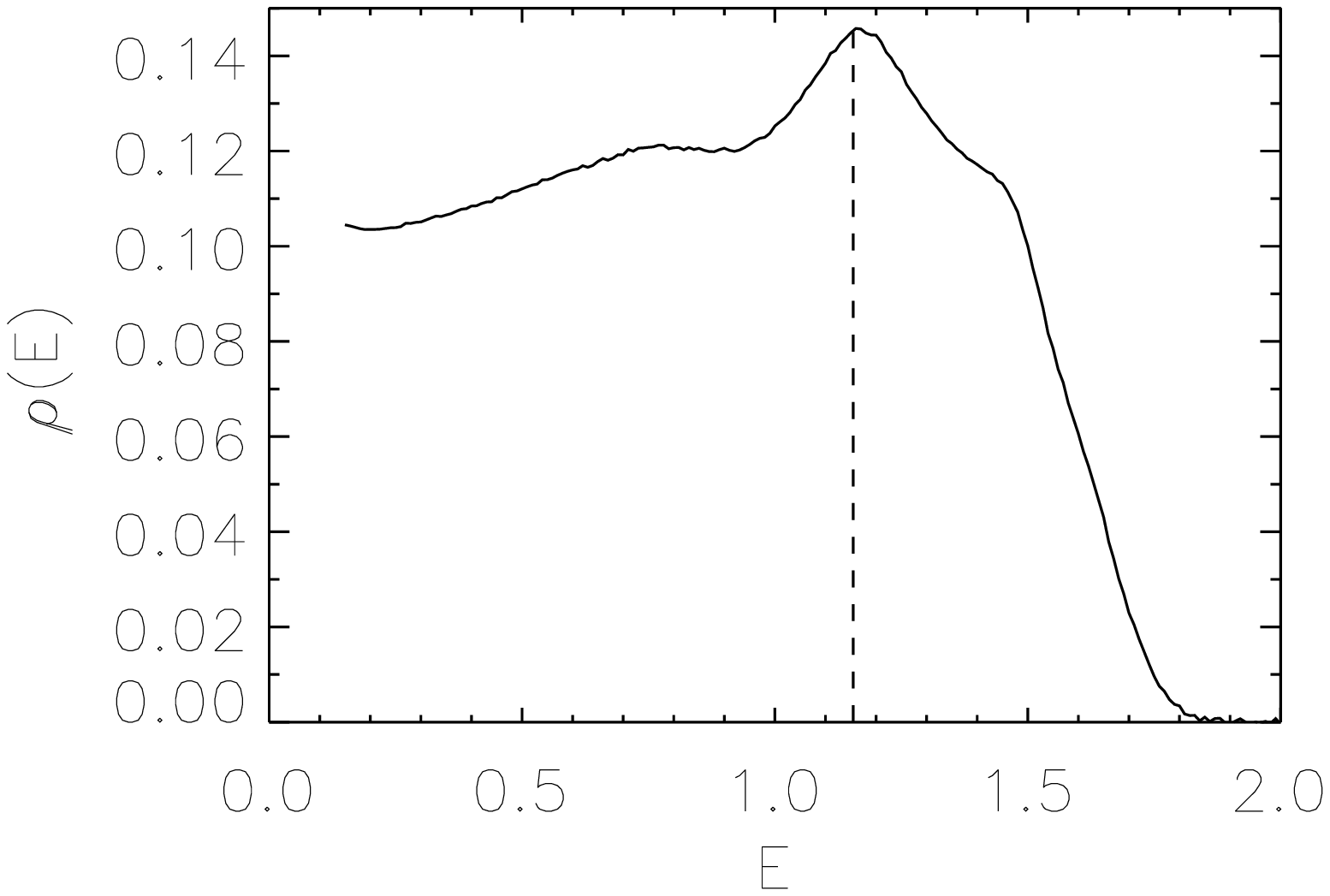}}
\end{center}
\protect\caption[5]{\label{fig:iter}Lyapunov exponent $\gamma(E)$ and
density of states $\rho(E)$ obtained by discretisation. 
The irregularities visible in Figure~\ref{fig:mc}.b have now disappeared.}
\end{figure}

Figure~\ref{fig:por} shows distributions $P(r,E)$ obtained that way. 
They are very similar to what was obtained by the MonteCarlo method,
cfr.~Figure~\ref{fig:rm}.b--c. In Figure~\ref{fig:iter} we display
the corresponding Lyapounov exponent $\gamma(E)$ and the density of
states $\rho(E)$.
The Lyapounov exponent is obtained simply by the discretised version
of Eq.~(\ref{mean-logr1}), and its derivatives are then
calculated by replacing them  by finite differences with a
differentiation interval $\Delta E = 0.025$.
(This value results from a compromise: For smaller $\Delta E$ a
complicated structure appears in the derivatives due to the
discretisation of $r$, whereas a too large $\Delta E$ smears out
details of $\rho(E)$.)

We see that both $\gamma(E)$ and $\rho(E)$ have essentially the same
shapes as we have previously encountered in figure 4, using
the Monte Carlo method. 
The irregularities which were visible in
Figure~\ref{fig:mc}.b have now disappeared in
Figure~\ref{fig:iter}.b. Moreover, the inset in 
Figure~\ref{fig:iter}.a, showing the asymptotic approach towards the
exact result (\ref{trivial}), indicates that the accuracy of
$\gamma(E)$ has been improved by roughly two orders of magnitude.

In the insets of Figures~\ref{fig:mc}.a and \ref{fig:iter}.a, the
difference $\gamma(E)- \log(E)$ seems to vanish at $E \simeq 1.8$. We
believe that this difference is non-zero up to 
$E=2$ but that it becomes very small ($\rho(E)$ has an essential singularity:
it vanishes 
 and all its derivatives 
vanish at $E=2$).  
The argument that eigenstates exist up to $E=2$
can be borrowed from the theory of Lifshitz tails \cite{Lif,Pastur2}
 in the Hermitian case:
in a random sample, one can find arbitrarily large regions where
$\theta_n \simeq \chi_n \simeq 0$.

\section{Perturbation theory}
\label{sec:perturb}

In this section we develop a perturbation theory in powers of $E$ to
determine the invariant distribution $P(r,E)$ solution of Eq.~(\ref{inv-dist}).
When this is known, we can obtain the corresponding perturbative
series for $\gamma(E)$ from Eq.~(\ref{mean-logr1}), and that of
$\rho(E)$ from Eq.~(\ref{rad-diff}).

For $E=0$ there are infinitely many solutions to
Eq.~(\ref{inv-dist}). Namely, the iteration formula (\ref{iter-r})
reduces to 
\beq
  r_{n+1} = \frac{1}{r_n},
\eeq
and any distribution preverving the symmetry $r \leftrightarrow 1/r$
(\ie such that $P(r,0)=P(1/r,0)/r^2$) is a solution. From
Eq.~(\ref{mean-logr1}) we thus immediately have
\beq
  \gamma(E=0) = 0.
\eeq

On the other hand, for $|E|>0$ the invariant distribution $P(r,E)$ is
unique, as witnessed by the numerical results of
Section~\ref{sec:numerics}. So the expansion in powers of $E$ that we
are going to develop presently is a degenerate perturbation theory, as
there are many solutions when $E=0$ and a single one, $P(r,E)$, when $E>0$.

It is useful to analyse first the small-$E$ expansion of the integral 
operator which appears in Eq.~(\ref{inv-dist}).
Consider two distributions $\tilde{Q}(r)$ and $Q(r)$ related by
\beq
\tilde{Q}(r) = \int_0^{2\pi} \frac{{\rm d}\varphi}{2\pi} \,
        \int_0^{\infty} {\rm d}s \,
        Q(s) \, \delta \left( r - \left| \frac{1}{s} +
        E {\rm e}^{{\rm i}\varphi} \right| \right).
 \label{operator}
\eeq

When $E<r$, this can be rewritten (by integrating over $s$)  as
\beq
\tilde{Q}(r) =
 \int_0^{2\pi} \frac{{\rm d}\varphi}{2\pi} \,
 {r^3 + r E^2 - 2 r E^2 \sin^2 \varphi + 2 r E  \cos{\varphi} \sqrt{r^2
 - E^2 \sin^2 \varphi} \over (r^2 - E^2)^2 \sqrt{r^2 - E^2 \sin^2
 \varphi}} 
        Q \left(
{E \cos \varphi + \sqrt{r^2 - E^2 \sin^2 \varphi} \over r^2 - E^2 }  \right).
 \label{operator1}
\eeq
Assuming $E \ll  r$ the integrand can then be expanded as a power
series in $E$, and after performing the integral over $\varphi$ we find
that
\bea
 \tilde{Q}(r) &=& \frac{1}{r^2} Q\left(\frac1r\right) +
          \frac{E^2}{4} \left[ \frac{9}{r^4} Q\left(\frac1r\right) +
          \frac{7}{r^5} Q'\left(\frac1r\right) +
          \frac{1}{r^6} Q''\left(\frac1r\right) \right] \nonumber \\
      &+& \frac{E^4}{64} \left[ \frac{225}{r^6} Q\left(\frac1r\right) +
          \frac{351}{r^7} Q'\left(\frac1r\right) +
          \frac{149}{r^8} Q''\left(\frac1r\right) +
          \frac{22}{r^9} Q^{(3)}\left(\frac1r\right) +
          \frac{1}{r^{10}} Q^{(4)}\left(\frac1r\right) \right] \nonumber \\
      &+& \frac{E^6}{2304} \left[ \frac{11025}{r^8} Q\left(\frac1r\right) +
          \frac{25839}{r^9} Q'\left(\frac1r\right) +
          \frac{18261}{r^{10}} Q''\left(\frac1r\right) +
          \frac{5382}{r^{11}} Q^{(3)}\left(\frac1r\right) +
          \right. \nonumber \\
      & & \left. \frac{732}{r^{12}} Q^{(4)}\left(\frac1r\right) +
          \frac{45}{r^{13}} Q^{(5)}\left(\frac1r\right) +
          \frac{1}{r^{14}} Q^{(6)}\left(\frac1r\right) \right]
          + {\cal O}(E^8).
 \label{expand}
\eea
We see that the effect of a non-zero energy is that $ \tilde{Q}(r)$ is not
just trivially related to $Q\left(\frac1r\right)$, but also depends on
all the derivatives of $Q$. Quite remarkably, the above
expansion can be written in  a compact form valid to all orders.
Defining the second order differential operator $M$  by
\begin{equation}
  M = \frac{{\rm d}}{{\rm d}r} \, r \, \frac{{\rm d}}{{\rm d}r} \, \frac1r, \\
\label{Mdef}
\end{equation}
the expansion of Eq.~(\ref{operator1}) to an arbirary order in $E$ can
be written
\begin{equation}
  \tilde{Q}(r)  = \sum_{n=0}^{\infty} \frac{E^{2n}}{ 4^n \ (n!)^2}
           M^n \left[ \frac{1}{r^2} Q\left(\frac1r\right)
           \right]. \label{P(r)} 
\end{equation}
It is straightforward to see that this 
 agrees with the na{\"\i}ve expansion formula (\ref{expand}). 
The validity of Eq.~(\ref{P(r)}) to an arbitrary order in $E$ 
is established in Appendix B.
Note that a simple change of variable $r \to 1/r$ transforms
Eq.~(\ref{P(r)}) into
\begin{equation}
  \tilde{Q}\left(\frac1r\right) =
           \sum_{n=0}^{\infty} { E^{2n} \over 4^n \ (n!)^2}
           L^n \left[ r^2 Q(r) \right] \label{P(1/r)},
\end{equation}
where the operator $L$ is defined by
\begin{equation}
  L = r^2 \, \frac{{\rm d}}{{\rm d}r} \, r \, \frac{{\rm d}}{{\rm d}r}\, r.
\label{Ldef}
\end{equation}

Coming back to the invariant distribution $P(r,E)$, we see by combining
Eqs.~(\ref{P(r)}) and (\ref{P(1/r)}) that it satisfies
\beq
 P(r,E) = \sum_{n=0}^{\infty} \left(E^2  \over 4 \right)^n \sum_{p=0}^n
  \frac{1}{[p! \, (n-p)!]^2} \, M^p \, \left[
  \frac{1}{r^2} L^{n-p} \left( r^2 P(r,E) \right) \right].
 \label{iter-twice}
\eeq
If we look for a solution $P(r,E)$ which can be expanded as
\beq
 P(r,E) = P_0(r) + {E^2 \over 4} P_1(r) + \left(E^2 \over 4 \right)^2
  P_2(r) + \cdots,
\eeq
we obtain by equating the two sides of Eq.~(\ref{iter-twice}), order
by order in $E$, that
\beq
 P_{m+1}(r) = \sum_{n=0}^{m+1} \sum_{p=0}^n \frac{1}{[p!\,(n-p)!]^2}\,
  M^p \, \left[ \frac{1}{r^2} L^{n-p} \left( r^2 P_{m+1-n}(r) \right) \right].
\eeq
The term with $n=0$ on the right-hand side is just $P_{m+1}(r)$, same as
the left-hand side, so that $P_{m+1}(r)$ disappears from the
equation. Therefore
\beq
 \sum_{n=1}^{m+1} \sum_{p=0}^n \frac{1}{[p!\,(n-p)!]^2}\,
 M^p \, \left[ \frac{1}{r^2} L^{n-p} \left(r^2 P_{m+1-n}(r)
 \right) \right] = 0.
\eeq
Moving all but the two $n=1$ terms, which are the only ones involving
$P_m(r)$, to the right-hand side we arrive at the equation
\bea
 M \left[  (1+r^4) P_m(r) \right]  &=& - M \sum_{n=2}^{m+1} \sum_{p=1}^n
      \frac{1}{[p!\,(n-p)!]^2} \, M^{p-1} \left[ \frac{1}{r^2} L^{n-p}
      \left( r^2 P_{m+1-n}(r) \right) \right] \nonumber \\
  &-& M \sum_{n=2}^{m+1} \frac{1}{[n!]^2} \, r^2 L^{n-1}
      \left( r^2 P_{m+1-n}(r) \right).
 \label{factorise}
\eea
Note that an $M$ operator has been factored out 
(by using the fact that $L r^2 = r^2 M r^4 $)
from all the
terms. Integrating over this, and shifting the summation variables,
we finally find
\bea
 (1+r^4) P_m(r) &=& A_m \, r \log r + B_m \, r -
     \sum_{n=1}^m \frac{1}{[(n+1)!]^2} \, r^2 \, L^n
     \left( r^2 P_{m-n}(r) \right) \nonumber \\
 &-& \sum_{n=1}^m \sum_{p=0}^n \frac{1}{[(p+1)! \, (n-p)!]^2} \,
     M^p \, \left[ \frac{1}{r^2} L^{n-p}
     \left( r^2 P_{m-n}(r) \right) \right],
 \label{all-orders}
\eea
where $A_m$ and $B_m$ are {\em a priori} arbitrary constants.

When $m=0$ we find
\beq
 P_0(r) = A_0 \, \frac{r \log r}{1 + r^4}
        + B_0 \, \frac{r}{1 + r^4}.
\eeq
By inserting this into Eq.~(\ref{P(r)}), with $Q(r)$ and
$\tilde{Q}(r)$ replaced by $P(r,E)$, we see that for $P_0(r)$ to
satisfy this equation to zeroth order in $E$, \ie
$P_0(r) = P_0(1/r)/r^2$, the constant $A_0$ should vanish,
whereas $B_0$ could be arbitrary (the problem is linear and $B_0$
would be fixed by the normalisation of $P(r,E)$).
This leads to $P_0(r) = B_0 r /(1+r^4)$.

For the same reason, namely that $P(r,E)$ is the fixed point of
Eq.~(\ref{P(r)}) rather than the solution of Eq.~(\ref{iter-twice})
obtained by by iterating Eq.~(\ref{P(r)}) twice,
one can show that $A_m = 0$ and $B_m$ is arbitrary for all $m$.
In fact, a non-zero $B_m$ produces a contribution to $P_m(r)$
proportional to $P_0(r)$, and so one might as well choose $B_m=0$ for
all $m \neq 0$ by allowing $B_0$ to depend on $E$.
In practise, then, all the $A_m=0$ for $m \geq 0$, all the $B_m = 0$
for $m >0$, and $B_0$ is an arbitrary function of $E$ (which can be
fixed by requiring that the distribution $P(r,E)$ be normalised).

In this way one arrives at
\beq
 P(r,E) = B_0(E) \left[ {r \over 1 + r^4}  + E^2 \left( {r^3 \over (1+
 r^4)^2} - {4 r^3 \over (1+ r^4)^4} \right) +  {\cal O}\big( E^{4}
 \big) \right].
\eeq

The recursion (\ref{all-orders}) gives an explicit formula for the
invariant distribution to {\em all orders} in  powers of $E$.
 We now show how to produce the 
equivalent expansion for the Lyapunov exponent.
First, it is easy  to check by induction  that  all the $P_m(r)$ are 
finite sums of  functions of the form
\beq
  f_n(r) = \frac{r}{(1+r^4)^n} \ \ \mbox{and} \ \
  g_n(r) = \frac{r^3}{(1+r^4)^n}.
\label{fndef}
\eeq
Indeed, this is true for $m=0$, and higher orders are generated from
Eq.~(\ref{all-orders}) by means of the operators $M$ and $L$. These
operators act on $f_n$ and $g_n$ as follows:
\bea
 M f_n &=& 16 n^2 g_{n+1} - 16 n (n+1) g_{n+2}, \nonumber \\
 M g_n &=& 4(1-2n)^2 f_n - 32 n^2 f_{n+1} +
              16 n(n+1) f_{n+2}, \nonumber \\
 L f_n &=& 4(1-2n)^2 g_n - 32 n^2 g_{n+1} +
              16 n(n+1) g_{n+2}, \\
 L g_n &=& 16 (1-n)^2 f_{n-1} - 16(1-3n+3n^2) f_n +
              48 n^2 f_{n+1} - 16 n(n+1) f_{n+2}. \nonumber
\eea
Applying Eq.~(\ref{all-orders}) repeatedly, we have found the
explicit expressions for the first seven orders of the (unnormalised)
invariant distribution. We report the first few orders here, and defer
the remaining ones to Appendix \ref{sec:high-orders}.
\bea
 P_0             &=& f_1, \nonumber \\
 \frac{P_1}{4}   &=& g_2 - 4 g_4, \nonumber \\
 \frac{P_2}{16}  &=& \frac{1}{2} f_2 + \frac{39}{2} f_3 - 22 f_4 -
   158 f_5 + 320 f_6 - 160 f_7, \nonumber \\
 \frac{P_3}{64}  &=& \frac{1}{2} g_2 - \frac{673}{18} g_3 -
   \frac{131}{18} g_4 + \frac{4456}{3} g_5 -
   \frac{842}{3} g_6 - 21840 g_7 \nonumber \\
   &+& 56520 g_8 - 53760 g_9 + 17920 g_{10}. \label{low-orders}
\eea
In order to compute the contributions to $\gamma(E)$ and to the
normalisation, one needs the following integrals
\bea
 I_n &=& \int_0^{\infty} {\rm d}r \, f_n(r), \nonumber \\
 J_n &=& \int_0^{\infty} {\rm d}r \, g_n(r), \nonumber \\
 K_n &=& \int_0^{\infty} {\rm d}r \, f_n(r) \log r, \nonumber \\
 L_n &=& \int_0^{\infty} {\rm d}r \, g_n(r) \log r.
\eea
Using (\ref{fndef}), these are easily evaluated
\bea
 I_{n+1} &=& \frac{2n-1}{2n} I_n \mbox{ with } I_1 = \frac{\pi}{4},
             \nonumber \\
 J_n     &=& \frac{1}{4(n-1)}, \nonumber \\
 K_{n+1} &=& \frac{2n-1}{2n} K_n - \frac{1}{4n} I_n
             \mbox{ with } K_1 = 0, \nonumber \\
 L_{n+1} &=& \frac{n-1}{n} L_n - \frac{1}{4n} J_n
             \mbox{ with } L_2 = 0.
\label{integrals}
\eea
Combining all of this we arrive at the expansion
\beq
 \gamma(E) = \frac{F(E)}{\pi G(E) + H(E)}
 \label{lyap-exp}
\eeq
with
\bea
 F(E) &=& \frac{1}{8} E^2 + \frac{5}{288} E^6 +
          \frac{43}{3600} E^{10} - \frac{2119}{235200} E^{14} +
          {\cal O}\big( E^{18} \big), \nonumber \\
 G(E) &=& \frac{1}{4} + \frac{9}{256} E^4 + \frac{8333}{294912} E^8 +
          \frac{2624621}{56623104} E^{12} +
          {\cal O}\big( E^{16} \big), \\
 H(E) &=& -\frac{1}{12} E^2 - \frac{89}{15120} E^6 +
          \frac{1088497}{29937600} E^{10} +
          \frac{576717747329}{490377888000} E^{14} +
          {\cal O}\big( E^{18} \big). \nonumber
\eea
It is worth noting that all the coefficients in the expansions of the
functions $F,G$ and $H$ are rational.
Equivalently,
\bea
 \gamma(E) &=& \frac{1}{2 \pi} E^2 + \frac{1}{6 \pi^2} E^4 +
   \left( \frac{1}{18 \pi^3} - \frac{1}{1152 \pi} \right) E^6 +
   \left( \frac{1}{54 \pi^4} - \frac{241}{20160 \pi^2} \right) E^8
   \nonumber \\
 &+& \left( \frac{1}{162 \pi^5} - \frac{2857}{362880 \pi^3} -
   \frac{31747}{3686400 \pi} \right) E^{10} +
   \left( \frac{1}{486 \pi^6} - \frac{1067}{272160 \pi^4} -
   \frac{11849219}{127733760 \pi^2} \right) E^{12} \nonumber \\
 &+& \left( \frac{1}{1458 \pi^7} - \frac{631}{362880 \pi^5} -
   \frac{9755539001}{160944537600 \pi^3} -
   \frac{530351263}{4161798144 \pi} \right) E^{14} \nonumber \\
 &+& \left( \frac{1}{4374 \pi^8} - \frac{709}{979776 \pi^6} -
   \frac{15094381}{502951680 \pi^4} -
   \frac{128124296107}{53137244160\pi ^2} \right) E^{16} +
   {\cal O}\big( E^{18} \big).
 \label{lyap-exp1}
\eea

\begin{figure}
\begin{center}
\leavevmode
\epsfysize=200pt{\epsffile{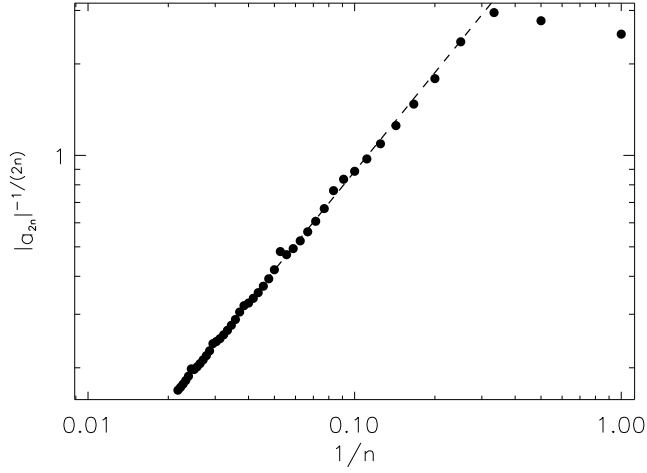}}
\end{center}
\protect\caption[5]{\label{fig:gam}Log-log plot of
$|a_{2n}|^{-1/(2n)}$ versus $1/n$.}
\end{figure}

In this way we can in principle generate $\gamma(E)$ to any order.
We calculated numerically the coefficients $a_{2n}$ of the expansion
of $\gamma(E)$
$$\gamma(E)= a_2 E^2 + a_4 E^4 + \cdots + a_{2n} E^{2n} + \cdots $$
up to $n= 46$ in order to estimate the radius of convergence of this
series. In Figure~\ref{fig:gam} we show a log-log plot of
$|a_{2n}|^{-1/(2n)}$ versus $1/n$. For $n>10$ the data are well fitted by
the form $|a_{2n}|^{-1/(2n)} \propto n^{-\nu}$ with $\nu \simeq 1.08$,
shown as a dashed line on the figure. This result indicates  
rather convincingly that the large-$n$ limit is zero, and thus
that the radius of convergence vanishes. We therefore believe that our
expansion is an asymptotic expansion.

\section{Discussion}
\label{sec:discussion}

In this paper we have determined numerically the Lyapounov exponent
$\gamma(E)$ and the density of states $\rho(E)$ of the one-dimensional
non-Hermitian Schr\"odinger equation (\ref{iter-psi}) when
the phases $\theta_n$ and $\chi_n$ are uniformly distributed. 
We have also developped a method of expanding these quantities in
powers of the energy $E$ and obtained $\gamma(E)$ up to order
$E^{16}$. Our procedure can be extended to obtain, in principle, an
arbitrarily high order.

The expansion in powers of $E$ is based on the expansion of the
integral equation (\ref{inv-dist}) satisfied by the invariant
distribution $P(r,E)$. Since this expansion is in principle only valid for
$E \ll r$, there is a possibility that the perturbation series for
$\gamma(E)$ suffers from the fact that when we integrate over $r$, we
use an expression valid for $r \gg E$ even in the neighborhood of
$r=0$.
 One cannot exclude for example  that when $r$ and $E$ become comparable
(and small), $P(r,E)$ becomes a scaling function of $r/E$.
This could invalidate our expoansion of $\gamma(E)$.
However
 when we compared our expansion with the results of the simulation
we found an excellent agreement for the first two terms and so it is
possible that our expansion, {\em a priori} valid for $r \gg E$,
remains valid down to $r=0$.

Our numerical results show that the density of states  vanishes
outside a circle of radius $|E|=2$, is non-zero inside this circle
$|E| < 2$, and has a non-trivial shape with a pronounced 
maximum at an energy $E= 2 / \sqrt{3}$.  The analyticity of
$\gamma(E)$ or of $\rho(E)$ at this energy remains an open problem.
Another interesting question would be to predict how $\rho(E)$ vanishes at $E=2$.

One could try to extend our approach to other situations, in
particular to cases where the phases $\theta_n$ and $\chi_n$ have a
non-uniform distribution \cite{FZ2}. In that case, the phase and the
modulus of the ratio $\psi_{n+1} / \psi_n$ are in general correlated
and one should look for the 
invariant distribution $P(r,E)$ of the {\it complex variables}
$r_n$ and $E$. Still, the fact that $\gamma(E=0)=0$ would remain true, and one
could try to expand the invariant distribution in powers of $E$
and use the Thouless formula to calculate $\rho(E)$.

\noindent{\large\bf Acknowledgments}

We would like to thank V.~Hakim and L.~Pastur for very useful and
interesting discussions.

\appendix

\section{Nature of the singularities of $P(r,E)$}
\label{app:sing}

In this Appendix we discuss the occurence of singularities in the
invariant measure $P(r,E)$  at the fixed point $r_+$  of the map 
$T_+(r) = E + \frac{1}{r}$, when one iterates the random map
(\ref{iter-r}) 
\begin{equation}
  r_{n+1}   = \left| \frac{1}{r_n} + E {\rm e}^{{\rm i} \varphi_n}
                \right|,
\label{rm1}
\end{equation}
with uniformly distributed $\varphi_n$ between $0$ and $2 \pi$.

We believe that the apparent singularities in $P(r,E)$ seen in
Figure~\ref{fig:rm} are due to events where several successive $r_n$
are close to the fixed point $r_+$, and that the nature of the
singularity can be understood by analysing only the neighbourhood of
$r_+$.

\begin{figure}
\begin{center}
\leavevmode
 \epsfysize=160pt{\epsffile{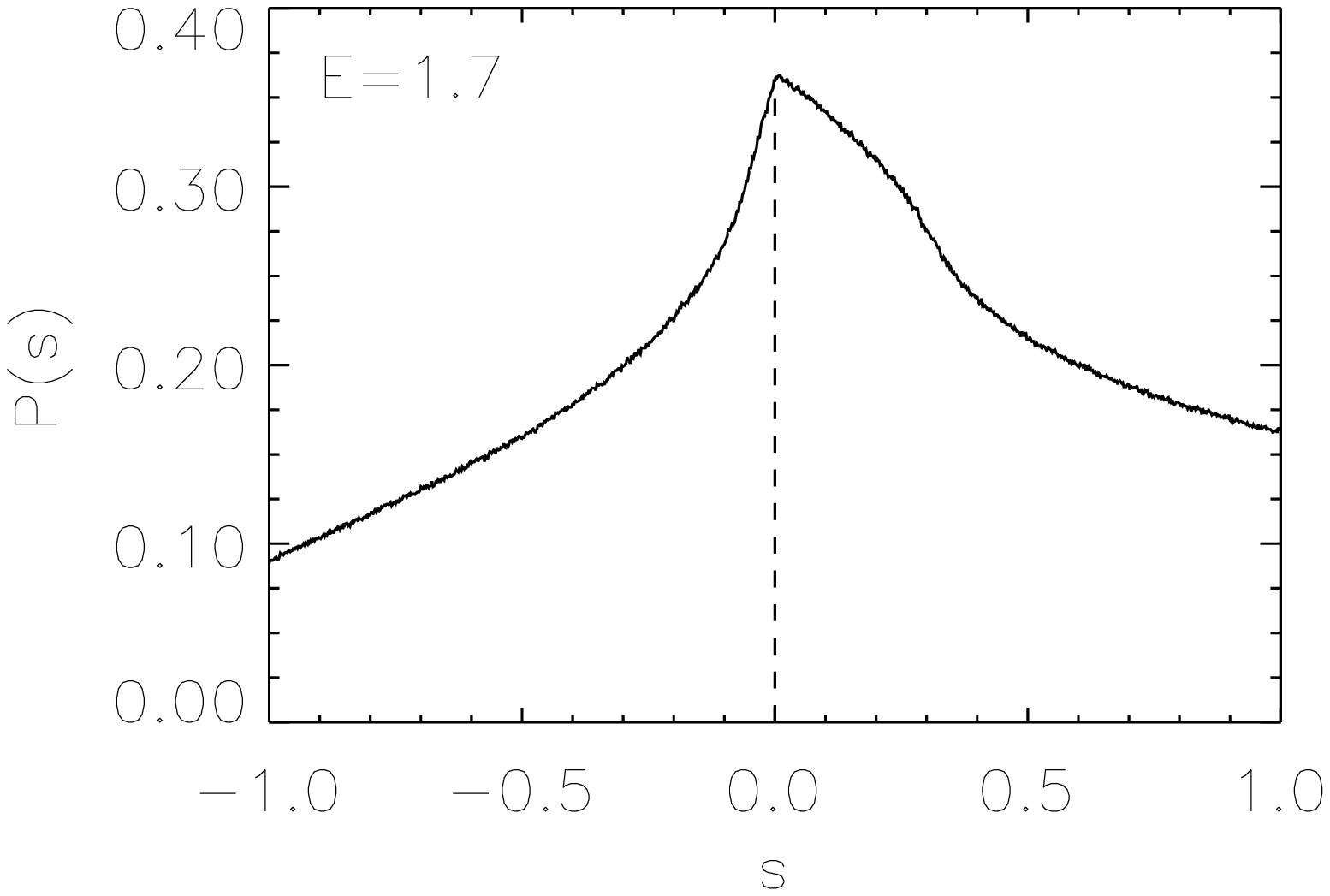}}
 \epsfysize=160pt{\epsffile{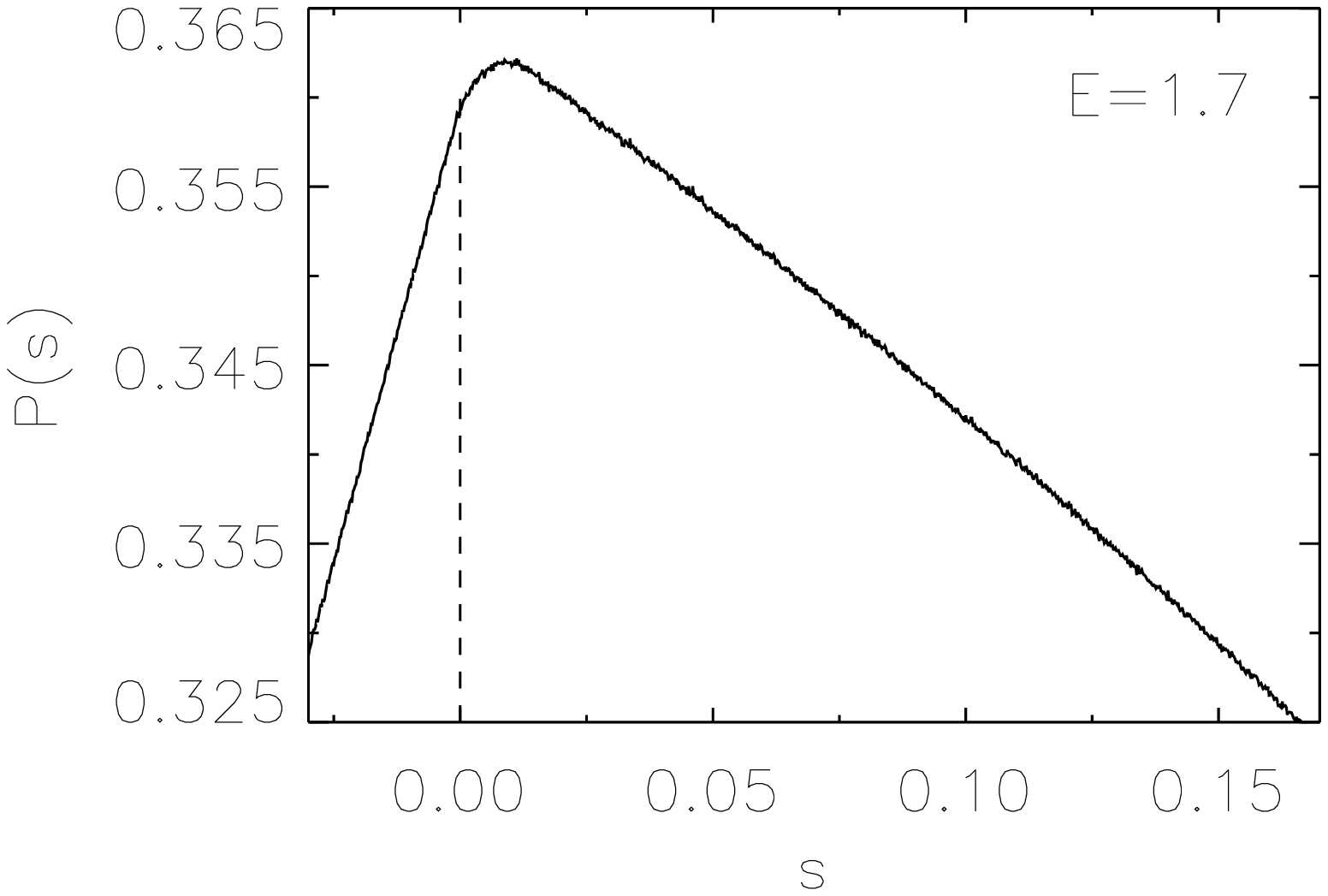}}
\end{center}
\protect\caption[5]{\label{fig:lin}Invariant distribution of $s_n$ for
the linearised problem defined by Eqs.~(\ref{rm2})--(\ref{rm2a}), here
with $E=1.7$. The magnification (obtained with $N=10^{11}$ iterations)
shows that there is in fact no singularity at $s=0$.}
\end{figure}

In physical terms, there is a competition between the deterministic
part of the map $T_+(r)$ which  tries to concentrate
the distribution on its attractive fixed point $r_+$, and
the noise due to $\varphi_n$ which tends to broaden the distribution.
To analyse the neighbourhood of $r_+$, we consider the linearised problem
\beq
 s_{n+1} = - a s_{n} + t_n,
 \label{rm2}
\eeq
where $a \in ]0,1[$ is a fixed slope, and $t_n$ is a random
positive number. The variable $s_n$ represents the difference $r_+ -
r_n$, when this difference is small. By expanding Eq.~(\ref{rm1}) to
second order in $\varphi_n$, it is seen that for Eqs.~(\ref{rm1}) and
(\ref{rm2}) to be equivalent in the neighbourhood of $r_+$, one should
choose 
\bea
 a   &=& - T_+'(r_+) = \frac{E^2 + 2 - E \sqrt{E^2 +4}}{2}, \nonumber \\
 t_n &=& \frac{E}{2 (E r_+ +1)} \  \varphi_n^2.
 \label{rm2a}
\eea
The essential feature of the distribution of $t_n$ is that
all the $t_n$ are positive and that the distribution  $Q(t)$ of $t_n$ 
diverges as $Q(t) \sim t^{-1/2} $ as $t \to 0$.
If we iterate (\ref{rm2}) numerically for $E=1.7$, that is for
$a=0.213851\cdots$, we recover at $s=0$ 
a singularity (see Figure~\ref{fig:lin}.a) which ressembles the one
seen on Figure~\ref{fig:rm}.c. Actually, the two distributions are
approximately mirror images near their respective fixed points, but
this is due to the fact that $s_n \simeq r_+ - r_n$
rather than $s_n \simeq r_n - r_+$.

Trying to analyse the nature of this singularity, we see from
Eq.~(\ref{rm2}) that $s_{n}$ can be written as 
\begin{equation}
 s_n = t_{n-1}  - at_{n-2} + a^2 t_{n-3} - a^3  t_{n-4} + a^4
 t_{n-5} - \ldots 
\label{rm3}
\end{equation}
or as 
\begin{equation}
 s_n =  y_n - a y'_n,
 \label{rm4}
\end{equation}
where $y_n$ and $y'_n$ are two positive independent and equally 
distributed random variables of the form
\bea
 y_n &=& \tau_n + a^2 \tau_{n-1} + a^4 \tau_{n-2} + a^6 \tau_{n-3} + \cdots,
 \nonumber \\
 y'_n &=& \tau'_n + a^2 \tau'_{n-1} + a^4 \tau'_{n-2} + a^6
 \tau'_{n-3} + \cdots.
 \label{yeq} 
\end{eqnarray}
All the $\tau_n$ and the $\tau'_n$ are independent and distributed
according to the same distribution $Q(t)$ as for the $t_n$. 

Let us first consider the case where $s_n$ is still given by
Eq.~(\ref{rm4}) but where the distribution $\pi(y)$ of the {\em positive}
variables $y$ is arbitrary. 
Using Eq.~(\ref{rm4}) one can show that the distribution $P(s)$ of $s$
is given by
\bea
 P(s) &=& \int_0^\infty {\rm d}y \, \pi(a y + s ) \  \pi(  y)
          \ \ { \rm for}  \ \  s >0 ,
 \label{pp} \\
 P(s) &=& \int_0^\infty {\rm d}y \, \pi(a y) \,
          \pi \left( y - \frac{s}{a} \right)
          \ \ { \rm for}  \ \  s <0.
 \label{pm}
\eea
One can calculate the successive derivatives with respect to $s$ at $s=0$,
and one obtains for the $n^{\rm th}$ derivative
\bea
 P^{(n)}(0_+) &=& \int_0^\infty {\rm d}y \, \pi^{(n)}(a y) \, \pi(  y),
 \label{dpp} \\
 P^{(n)}(0_-) &=& \left( {-1 \over a} \right)^n \int_0^\infty {\rm d}y
  \, \pi(a y) \, \pi^{(n)}(y).
\label{dpm} 
\end{eqnarray}
If we assume that $\pi(y)$ and its first
$\left[ \frac{n+1}{2} \right]$ derivatives vanish at the origin, and
as $y \to \infty$,
it is easy to see by integration by parts that Eqs.~(\ref{dpp}) and
(\ref{dpm}) coincide, so that $P(s)$ has at least $n$ derivatives at $0$.

We are now going to argue that since $y_n$ is given by Eq.~(\ref{yeq}),
its stationary  distribution $\pi(y)$  vanishes at $y=0$ as well 
as all its derivatives. To see this, let us consider the generating
function $\langle {\rm e}^{- \beta y} \rangle $ of $y$. From
Eq.~(\ref{yeq}) one has 
\beq
 \langle {\rm e}^{- \beta y} \rangle = \prod_{n=0}^\infty
 \left[\int_0^\infty {\rm d}t \, Q(t) \,  {\rm e}^{ - \beta t a^{2n}} \right].
\eeq
Because $Q(t) \sim t^{-1/2}$ for small $t$, one has for large $\beta$
\beq
 \int_0^\infty {\rm d}t \, Q(t) \, {\rm e}^{- \beta t}
 \simeq \frac{B}{\beta^{1/2}}.
\eeq
For large $\beta$, one has $\langle e^{-\beta y} \rangle \simeq 
\langle e^{-\beta a^2 y} \rangle $ and  one finds that to leading order 
\beq
 \log \left[ \langle {\rm e}^{- \beta y} \rangle \right]  \simeq
 \frac{1}{8} \frac{\log^2 \beta}{\log a}.
\eeq
We see that $\langle e^{ - \beta y} \rangle$ vanishes faster 
(as $\log a < 0$)  than any
negative power of $\beta$ as $\beta \to \infty$.
As a consequence $\pi(y)$ and all its derivatives vanish at $y=0$.

Therefore we can conclude that all the derivatives of the distribution
$P(s)$ are continuous at $s=0$. This is confirmed by a magnification
of the small-$s$ region, as seen in Figure~\ref{fig:lin}.b,
where the rounding of $P(s)$ at $s=0$ becomes visible.

Coming back to the stationary distribution $P(r,E)$, we observe the
same rounding of the apparent singularity at $r_+$ (see
Figure~\ref{fig:rm17}).

\section{Justification of Eq.~(\ref{P(r)})}
\label{app:P(r)}

Imagine that $s$ is a random positive variable distributed according to some
given distribution $Q(s)$ and let $r$ be given by
\beq
 r = \left| {1 \over s} + E e^{{\rm i} \varphi} \right|,
 \label{r(s)}
\eeq
where $\varphi$ is uniformly distributed between $0$ and $2 \pi$.
We want to show that the distribution $\tilde{Q}(r)$ of $r$ can be written as
in Eq.~(\ref{P(r)}): 
\beq
\tilde{Q}(r) = \sum_{n=0}^{\infty} \frac{E^{2n}}{ 4^n \ (n!)^2} 
           M^n \left[ \frac{1}{r^2} Q\left(\frac1r\right)
           \right],
 \label{PP(r)}
\eeq
where 
\beq
 M = \frac{{\rm d}}{{\rm d}r} \, r \, \frac{{\rm d}}{{\rm d}r} \, \frac1r
\eeq

Let us assume for simplicity that all (positive and negative) moments
of $Q(s)$ exist. By taking the $(2p)^{\rm th}$ power of
Eq.~(\ref{r(s)}) and by averaging over $\varphi$, one can see that
\beq
 \int_0^\infty {\rm d}r \, r^{2p} \tilde{Q}(r) =
 \sum_{n=0}^{p} \frac{[p!]^2}{[n!]^2 [(p-n)!]^2} E^{2n}
 \int_0^\infty {\rm d}s \, \frac{1}{s^{2p-2n}} Q(s).
 \label{PP1(r)}
\eeq
We see that for Eqs.~(\ref{PP(r)}) and (\ref{PP1(r)}) to be
equivalent, one simply need to show that
\beq
 \frac{1}{4^n} \int_{0}^\infty {\rm d}r \, r^{2p} \,
 M^n \left[ \frac{1}{r^2} Q\left(\frac1r\right) \right] = 
 \frac{[p!]^2}{[(p-n)!]^2} 
 \int_0^\infty {\rm d}s \, \frac{1}{s^{2p-2n}} Q(s).
\eeq
For $n=0$ this is an obvious identity, and for $n>0$ it can be
established by induction, using integrations by parts.

\section{Higher orders of $P(r,E)$}
\label{sec:high-orders}

Here we list the higher order terms of $P(r,E)$, obtained from
Eq.~(\ref{all-orders}),
expressed in terms of the elementary functions $f_n$ and $g_n$. These
higher order terms are used together with Eq.~(\ref{low-orders}) to
obtain the expansion (\ref{lyap-exp}) for the Lyapunov exponents.
\bea
 \frac{P_4}{4^4} &=& \frac{221}{8} f_2 + \frac{16295}{72} f_3 -
   \frac{94193}{18} f_4 - 25803 f_5 + \frac{2529101}{9} f_6 +
   \frac{2619745}{9} f_7 \nonumber \\
   &-& \frac{22135892}{3} f_8 + \frac{77875796}{3} f_9 - 43988480 f_{10} +
   40636800 f_{11} - 19712000 f_{12} \nonumber \\
   &+& 3942400 f_{13}, \\
 \frac{P_5}{4^5} &=& -\frac{413}{72} g_2 - \frac{370801}{648} g_3 +
   \frac{11887541}{3240} g_4 + \frac{34137437}{135} g_5 -
   \frac{76908091}{135} g_6 \nonumber \\
   &-& \frac{174769499}{9} g_7 + \frac{684150977}{9} g_8 +
   \frac{1394447776}{3} g_9 - \frac{38774321284}{9} g_{10} \nonumber \\
   &+& 14632026816 g_{11} - 27725662640 g_{12} + 31886412800 g_{13} -
   22180787200 g_{14} \nonumber \\
   &+& 8610201600 g_{15} - 1435033600 g_{16}, \\
 \frac{P_6}{4^6} &=& \frac{132619}{288} f_2 + \frac{55698437}{2592} f_3 -
   \frac{1031521993}{1080} f_4 - \frac{10179687761}{1080} f_5 +
   \frac{76720899118}{405} f_6 \nonumber \\
   &+& \frac{166044285683}{270} f_7 - \frac{2068679706428}{135} f_8 +
   \frac{3096140849561}{135} f_9 + \frac{1697546657546}{3} f_{10} \nonumber \\
   &-& 4380421880262 f_{11} + \frac{147448995026264}{9} f_{12} -
   38217908241192 f_{13} \nonumber \\
   &+& \frac{179597752851712}{3} f_{14} - \frac{192990668440960}{3} f_{15} +
   46970084761600 f_{16} \nonumber \\
   &-& 22329532825600 f_{17} + 6245266227200 f_{18} -
    780658278400 f_{19} \\
 \frac{P_7}{4^7} &=& -\frac{288203}{2592} g_2 -
   \frac{1354843111}{23328} g_3 + \frac{815716196107}{583200} g_4 +
   \frac{16419260711323}{170100} g_5 \nonumber \\
   &-& \frac{207216071509033}{340200} g_6 - \frac{113911389465407}{5670} g_7 +
   \frac{579469587812377}{5670} g_8 \nonumber \\
   &+& \frac{677054300749364}{405} g_9 - \frac{1078075200453799}{81} g_{10} -
   \frac{324458646218758}{9} g_{11} \nonumber \\
   &+& \frac{8114117353035130}{9} g_{12} -
   \frac{50273274830679712}{9} g_{13} + \frac{180446654936149976}{9} g_{14}
   \nonumber \\
   &-& \frac{433731592050615872}{9} g_{15} +
   \frac{735643743703812320}{9} g_{16} - \frac{298906446014525440}{3} g_{17}
   \nonumber \\
   &+& \frac{261230176431411200}{3} g_{18} - 53361562052198400 g_{19} +
   21803255983308800 g_{20} \nonumber \\
   &-& 5339702624256000 g_{21} + 593300291584000 g_{22}.
\eea

\end{document}